 \documentclass[sigconf]{acmart}  

\usepackage[ruled,vlined,linesnumbered]{algorithm2e}
\usepackage{amsmath}

\usepackage{multicol, multirow}
\usepackage{color}
\usepackage{graphicx}
\usepackage{rotating}
\usepackage{pifont}
\usepackage{array}
\usepackage{makecell}
\usepackage{tabularx,booktabs} 
\definecolor{munsell}{rgb}{0.13, 0.55, 0.13}
\newcommand{\cmark}{\color{munsell}\ding{51}}%
\newcommand{\xmark}{\color{red}\ding{55}}%

\usepackage{marginnote}
\usepackage{cleveref}
\usepackage{caption}
\usepackage{subcaption}

\AtBeginDocument{%
  \providecommand\BibTeX{{%
    \normalfont B\kern-0.5em{\scshape i\kern-0.25em b}\kern-0.8em\TeX}}}

\newcommand{\name}{\textsc{ControlGUI}\xspace}

\acmSubmissionID{}

\begin{document}


\title[\name: Multimodal GUI Generation for Rapid Early-Stage Exploration with Dual-Adapter Diffusion Model]{\name: Rapid GUI Exploration with Multimodal Dual-Adapter Diffusion Model}

\author{Aryan Garg}
\email{aryan.garg@aalto.fi}
\authornote{Contributed equally.}
\affiliation{
    \institution{Aalto University} 
    \country{Finland}
}

\author{Yue Jiang}
\email{yue.jiang@aalto.fi}
\authornotemark[1]
\affiliation{
    \institution{Aalto University} 
    \country{Finland}
}

\author{Antti Oulasvirta}
\email{antti.oulasvirta@aalto.fi}
\affiliation{
    \institution{Aalto University} 
    \country{Finland}
}








\renewcommand{\shortauthors}{Garg et al.}

\begin{abstract}          
Exploring diverse GUI ideas early in the design process is challenging: existing tools either constrain creativity or demand overly detailed inputs.
Generative AI could help, yet current systems rely mostly on text prompts and rarely preserve designers’ intended layouts or attention patterns.
We present \name, a multimodal generative system that supports GUI exploration through text, wireframes, and visual flow specifications.
By combining these inputs flexibly, \name generates large galleries of low-fidelity GUI concepts that align with both structural layouts and intended user attention.
Our contributions include a dual-adapter diffusion model that jointly controls wireframe and visual flow, and a large-scale multimodal dataset of 72,500 GUIs with aligned screenshots, wireframes, descriptions, and scanpaths.
Through qualitative and quantitative evaluations, along with a user study simulating early-stage ideation, we show that multimodal control enhances perceived diversity, creativity, and alignment with designers’ intent, providing a flexible and effective tool for rapid GUI concept exploration.

\end{abstract}

\newcommand{\loss}{\mathcal{L}}
\newcommand{\image}{\mathcal{I}}
\newcommand{\encoder}{\mathcal{E}}
\newcommand{\decoder}{\mathcal{D}}
\newcommand{\normal}{\mathcal{N}}





\keywords{User Interface, User Interface Design, GUI Design Exploration, Visual Flow, Diffusion Model, Multi-modal Synthesis, Large Language Models, Vision Language Models}


\begin{teaserfigure}
  \def\w{\linewidth}
  \centering
  \includegraphics[width=\w]{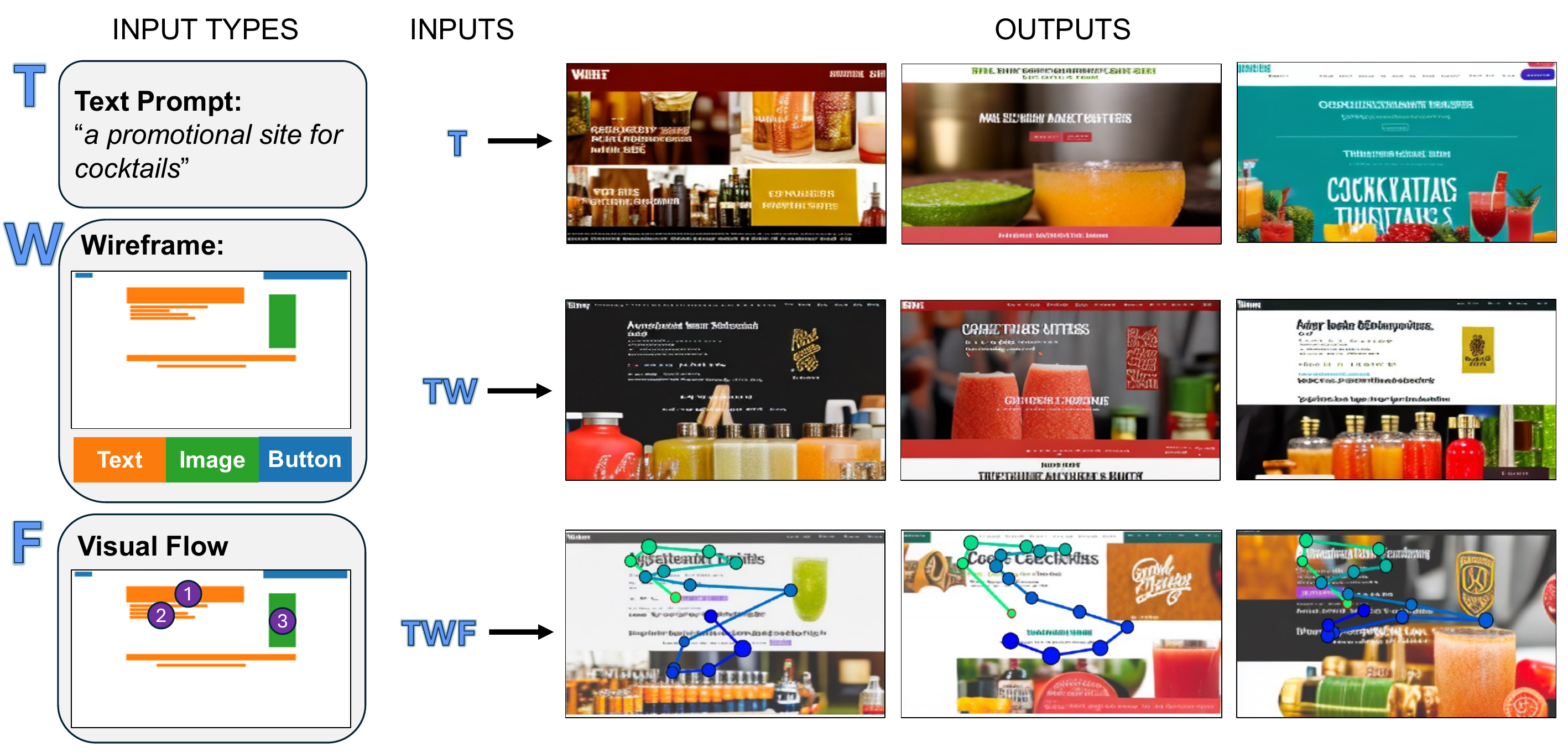}
  \caption{
Given different combinations of input modalities, T (text prompt), W (wireframe), and F (visual flow), our model generates GUIs that are both diverse and aligned with the specified conditions. With only a text prompt (T), it produces stylistically varied interfaces matching the description. Adding a wireframe (TW) enforces structural fidelity, ensuring the correct placement of elements such as text, images, and buttons. Incorporating visual flow (TWF) further constrains the design by guiding user attention across elements in the intended sequence, providing perceptual control over the interface. The scanpaths over generated GUIs are visualized with a color gradient (green → blue) showing temporal progression, with fixation points marked as circles, which closely match the specified element order.
  }
  \Description{This figure shows that given different combinations of input modalities, text prompt, wireframe, and visual flow, our model generates GUIs that are both diverse and aligned with the specified conditions. With only a text prompt, it produces stylistically varied interfaces matching the description. Adding a wireframe  enforces structural fidelity, ensuring the correct placement of elements such as text, images, and buttons. Incorporating visual flow further constrains the design by guiding user attention across elements in the intended sequence, providing perceptual control over the interface.}
  \label{fig:teaser}
\end{teaserfigure}


\maketitle

\section{Introduction}

In the early stages of graphical user interface (GUI) design, designers need to explore a large number of alternatives. 
Doing so allows them to generate ideas, compare trade-offs, and spark creative solutions to complex design problems~\cite{dow2011prototyping, tohidi2006getting}. 
To avoid premature commitment, designers often use low-fidelity representations such as sketches, wireframes, or intentionally underspecified mockups~\cite{landay1996silk}. 
These lightweight representations let designers explore the design space without investing in high-fidelity details before key conceptual directions are established. 
However, producing many such variations is challenging in practice: designers face time constraints, risk fixation on early ideas~\cite{jansson1991design}, and often lack tools that allow them to express vague or partial ideas without requiring over-specification.

Although design exploration tools have received significant attention in HCI research, supporting the critical early stage of GUI design remains unresolved. 
Sketching-based systems reduce the effort of creating digital variants by capturing and refining pen-and-paper sketches~\cite{landay1996silk}. 
Search and retrieval tools provide inspiration by surfacing existing GUIs~\cite{herring2009getting, kumar2013webzeitgeist, li2021screen2vec}, and constraint-based systems allow designers to generate alternatives within user-specified boundaries~\cite{swearngin2020scout, jiang2019orclayout}. 
However, these approaches fall short in two key ways: they either limit designers’ creative freedom by anchoring ideas to initial sketches or existing examples, or they demand detailed specifications that designers are unwilling or unable to provide early on. 
As a result, there is no method that allows designers to rapidly generate a wide range of diverse, low-fidelity GUI concepts while retaining the flexibility to work with partial, underspecified, or evolving ideas.

Recent advances in generative AI hold promise for addressing this gap, but current systems remain poorly matched to early-stage GUI design needs. 
Most rely exclusively on text prompts, whereas designers rarely conceive interface ideas as fully articulated descriptions. 
Instead, they naturally think in mixed modalities: a rough textual idea here, a quick sketch of layout there, or an envisioned visual flow of user attention~\cite{landay1996silk}. 
Existing systems also struggle to preserve designers’ intent in both structural and perceptual aspects, providing little support for perceptual goals such as visual flow, the sequence in which users attend to interface elements, which plays a crucial role in usability~\cite{Rosenholtz11, Still10, ueyes} but is rarely considered early due to the cost of empirical measurement.

We hypothesize that enabling multimodal control is key to making generative AI truly useful for early-stage GUI exploration. 
Designers should be able to guide generation using whichever medium feels most natural at a given moment, text, sketch, or perceptual specification, and combine these inputs flexibly without being forced into a rigid format. 
Supporting multiple modalities simultaneously allows a system to better fit creative workflows and preserve both structural and perceptual intent.

In this paper, we present \name, a controllable diffusion model for early-stage GUI generation that supports flexible multimodal input. From a designer’s perspective, \name lets designers rapidly explore interface ideas by generating galleries of low-fidelity GUI concepts within minutes. 
Designers can guide the system using any combination of three optional specifications: (a) natural language prompts, (b) wireframes describing layout and element types, and (c) visual flows indicating intended sequences of user attention (e.g., the order of elements), without being forced to fully define every detail. This approach lets designers experiment with low effort and uncover promising directions before committing to high-fidelity designs.

Our key technical insight is that effective GUI generation requires simultaneous control over both layout and perceptual flow, something existing systems fail to achieve. 
To realize this, we design a diffusion model with a dual-adapter architecture built on Stable Diffusion~\cite{rombach2022high}.
One adapter provides fine-grained control over element placement and layout, while the other guides higher-level perceptual goals, such as visual flow. 
This dual mechanism enables the model to generate interfaces that are coherent not only in structure but also in how users are likely to perceive them.

To evaluate \name, we adopted a two-pronged approach. First, qualitative and quantitative comparisons show that our method better aligns with designer intent than text-only or layout-only baselines and can generate 100 diverse designs in under two minutes. Second, a user study simulating early-stage GUI ideation further demonstrates that multimodal control improves perceived diversity, creativity, and alignment with intent. Participants highlighted \name’s value as a flexible ideation assistant for exploring diverse GUI concepts rapidly.

\noindent In summary, this paper makes the following contributions:
\begin{enumerate}
 \item A multimodal control method for GUI diffusion models. We introduce \name, a controllable diffusion model that supports text, wireframe, and visual flow specifications as flexible, combinable inputs. This allows designers to rapidly explore diverse, low-fidelity GUI concepts without requiring fully specified prompts.
 \item A dual-adapter diffusion architecture. we introduce complementary adapters, one for wireframe and one for visual flow, that together allow designers to guide both structural and perceptual aspects of GUI generation, producing concepts that are aligned with design intent.
 \item A large-scale multimodal GUI dataset. We contribute a dataset of 72,500 mobile GUIs and webpages, including aligned screenshots, wireframes, textual descriptions, and user scanpaths. This supports training and evaluation of multimodal generative models that reason jointly about structure, semantics, and visual flow.
\end{enumerate}

We will open-source our code and dataset after publishing.


\section{Related Work}

This section reviews key challenges in GUI generation and recent advances in diffusion models.

\begin{table}[t]
    \centering
       \caption{Comparison of GUI generation models: Our model advances the field by allowing flexible inputs and diverse GUI outputs without requiring manual, detailed specifications, such as defining constraints.
}
      \resizebox{\linewidth}{!}{
    \begin{tabular}{lm{2cm}m{2cm}m{1.8cm}m{1.5cm}m{1.87cm}m{1.7cm}}
    \toprule
         \textbf{Model} & \textbf{Application}  & \textbf{Inputs} & \textbf{Flexible Multimodal Input?}  &  \textbf{Generative Model?}  &  \textbf{No Manual Detailed Specification Required?} &  \textbf{Complete GUI Presentation?} \\
    \midrule
         Webzeitgeist~\cite{kumar2013webzeitgeist} & GUI Retrieval & Wireframe & \xmark  & \xmark  & \cmark & \cmark  \\
         Scout~\cite{swearngin2020scout}  & Constraint-based Optimization & GUI Elements, Constraints & \xmark & \xmark  & \xmark & \cmark\\
          UICoder~\cite{wu2024uicoder} & GUI Code Generation  & Prompt & \xmark  & \cmark & \cmark & \cmark\\
          PLay~\cite{cheng2023play} & Layout Generation  & Gridlines & \xmark  & \cmark  & \cmark & \xmark\\
          CoLay~\cite{cheng2024colay} & Layout Generation & Prompt, Gridlines, Element Type Count  & \cmark & \cmark & \cmark & \xmark\ \\
         \midrule
         \bf Ours  & GUI Exploration & Prompt, Wireframe, Visual Flow & \cmark & \cmark & \cmark & \cmark \\        
    \bottomrule
    \end{tabular}
    }

\Description{This table shows the comparison of GUI generation models. Our model advances the field by allowing flexible inputs and diverse GUI outputs without requiring manual, detailed specifications, such as defining constraints.}
    \label{tab:table_model}
\end{table}

\subsection{GUI Generation}

Exploring multiple GUI alternatives is essential for effective GUI design, as comparing diverse options enables stronger critique and more informed decision-making~\cite{dow2011prototyping, tohidi2006getting, jiang2022computational, jiang2024computational2, jiang2023future, jiang2024computational}. \autoref{tab:table_model} shows a comparison of representative GUI generation methods.

Generative methods for GUIs should support flexible multimodal inputs to support open-ended exploration. Although design literature often advocates for the use of rough sketches to explore design ideas~\cite{boyarski1994computers, landay1996silk}, sketching alone can be limiting. 
Designers may face challenges such as fixation, which can constrain the generation of novel solutions~\cite{jansson1991design}. 
Previous constraint systems have facilitated the exploration of alternatives through the application of constraints~\cite{swearngin2020scout, jiang2019orclayout, jiang2020orcsolver, jiang2020reverseorc, jiang2024flexdoc}. 
However, these systems typically require detailed manual specifications, which are cumbersome and time-consuming in the early design phase.

It is important to generate diverse, realistic GUI ideas that align with the designers intent. 
GUI retrieval techniques have been employed to obtain exemplar GUIs~\cite{herring2009getting, kumar2013webzeitgeist, li2021screen2vec}. However, the retrieved designs are often bound by the limitations of their source databases and may not meet the specific needs of the current design task. Recent advances in generative models have broadened the scope of exploration of design alternatives, but they still face some challenges. Some methods only generated layouts without visual graphics, making it harder to envision final designs~\cite{cheng2023play, cheng2024colay}. Other methods used text prompts to generate GUIs~\cite{wu2024uicoder}. However, text-based prompts often fall short in capturing certain GUI characteristics effectively, which are better conveyed through visual means, and frequently produce oversimplified GUIs. Moreover, none of these approaches consider user attention, a important but often overlooked factor in GUI design.

Additionally, designers also aim to shape how users perceive a GUI design, not just how it looks. 
An important aspect of this user experience is \emph{visual flow}: the sequence in which users visually navigate the GUI.  
Effective visual flow can improve engagement, support task completion, and guide attention to key elements~\cite{Rosenholtz11, Still10, ueyes, jiang2024ueyes, wang2024visrecall++, emami2024impact}. 
Thus, designers often care about how GUIs guide users' attention on the GUIs~\cite{Rosenholtz11, Still10, ueyes}. 
However, considering visual flow typically requires empirical user studies, which are resource-intensive and time-consuming.
Such evaluations often involve recruiting participants to interact with prototypes after each iteration, making them impractical for early-stage ideation.
Consequently, visual flow is rarely considered early in the design process.
Furthermore, existing generative approaches do not support specifying such attention-guiding goals.
Designers’ intentions around visual flow are difficult to express through current prompting interfaces and remain unsupported in current GUI generation tools.

To address these gaps, we present the first diffusion-based generative model for controllable GUI exploration that integrates visual flow into the generation process.
Our model accepts any combination of three optional inputs, natural language prompt, wireframe, and visual flow, allowing designers to work in the modality or level of detail that best matches their stage of ideation.
Built on Stable Diffusion~\cite{rombach2022high}, our approach produces complete GUI prototypes including visual styling, graphical elements, and automatically inpainted meaningful text.
This enables rapid generation of hundreds of diverse, realistic concepts in seconds, supporting broad, efficient exploration of the design space while more closely aligning with designer intent.

\subsection{Diffusion Models}

Given the limitations of current GUI generation approaches, recent advances in diffusion models offer a potentially promising way to support flexible inputs and diverse outputs for GUI design.
Recent advances in diffusion models have achieved substantial improvements in image quality, as seen in models such as GLIDE~\cite{nichol2021glide}, DALL-E2~\cite{ramesh2022hierarchical}, Imagen~\cite{saharia2022photorealistic}, and Latent Diffusion Models~\cite{rombach2022high}. These approaches provide stable training and support multi-modal conditional generation without requiring separate objectives for each modality.
In particular, Stable Diffusion~\cite{rombach2022high}, a widely used implementation of Latent Diffusion Models, is designed to efficiently generate high-quality images while accommodating a wide range of conditional inputs beyond text prompts, such as bounding boxes or structural layouts. 
Further improvements in guidance techniques, including classifier guidance~\cite{dhariwal2021diffusion} and classifier-free guidance~\cite{ho2022classifier}, are applied to enhance the ability of these models to align outputs closely with input specifications.
In our work, we build on Stable Diffusion with classifier-free guidance to produce controllable GUI generations. This combination offers both the flexibility to incorporate varied input modalities and the precision needed to match the design intent, making it well aligned with the requirements of early-stage GUI exploration.

\paragraph{Adapters in Diffusion Models}  

To enhance controllability in the diffusion model, we employ adapter techniques.
Adapters~\cite{adapters_in_nlp} are lightweight neural modules that can modify, specialize, or extend pretrained models without requiring full retraining. 
In diffusion models, adapters serve as modular components that inject additional conditioning signals, enabling the model to respond to domain-specific requirements while preserving its general capabilities.
By attaching such modules to the diffusion models, new control mechanisms and input modalities can be enabled with minimal computational cost.

Recent works illustrate the effectiveness of this approach. ControlNet~\cite{controlnet} applies the Side-Tuning~\cite{side_tune} approach in each block, adding zero-convolutions to integrate conditional information via a trainable copy of the Stable Diffusion encoder.
Similarly, IP-Adapter~\cite{ye2023ip} modifies cross-attention layers and introduces condition-specific encoders to accommodate new input modalities. T2I-Adapter~\cite{mou2024t2i} adopts a side-convolutional network for blending outputs layer by layer, also following the Side-Tuning principle.
Building on these ideas, we propose a dual-adapter configuration that provides simultaneous structural and perceptual control: one adapter targets local properties, such as the positioning and type of GUI elements, while the other governs global properties, such as the intended visual flow across the interface.

\begin{table}[t]
    \centering
      \caption{Comparison of GUI datasets: Our dataset uniquely combines mobile and webpage GUIs with comprehensive annotations and predicted scanpaths.
}
      \resizebox{\linewidth}{!}{
    \begin{tabular}{lccccc}
    \toprule
         \textbf{Dataset} & \textbf{GUI Types} & \textbf{$\#$ GUIs} & \textbf{Detections} & \textbf{Prompts} & \textbf{Scanpaths} \\
    \midrule
         MUD~\cite{mud} & Mobile (Android) & $\sim$18,000 & \cmark & \xmark & \xmark \\  
         RICO~\cite{rico} & Mobile (Android) & $\sim$72,000 & \cmark & \xmark & \xmark \\
         RICO-Semantic~\cite{rico_semantic} & Mobile (Android) & $\sim$72,000 & \cmark & \xmark & \xmark \\
         Clay~\cite{li2022learning} & Mobile (Android) & $\sim$60,000 & \cmark & \xmark & \xmark \\
          ENRICO~\cite{enrico} & Mobile (Android) & 1,460 & \cmark & \xmark & \xmark \\ 
          VINS~\cite{vins} & Mobile (Android + IOS) & 4,543 & \cmark & \xmark & \xmark \\ 
          AMP~\cite{zhang2021screen} & Mobile  (IOS) (closed source) & $\sim$77,000 & \cmark & \xmark & \xmark \\ 
          Webzeitgeist~\cite{kumar2013webzeitgeist} & Webpages & 103,744 & \cmark & \xmark & \xmark \\
         WebUI~\cite{webui} & Webpages & $\sim$350,000 & \cmark & \xmark & \xmark \\  
         UEyes~\cite{ueyes} & Mobile, Webpages, Poster, Desktop & 1,980 & \xmark & \xmark & \cmark \\
         \midrule
         \bf Ours & Webpages, Mobile (Android + IOS) & $\sim$72,500 & \cmark & \cmark & \textcolor{munsell}{Predicted} \\        
    \bottomrule
    \end{tabular}
    }
\Description{This table shows the comparison of GUI datasets: Our dataset uniquely combines mobile and webpage GUIs with comprehensive annotations and predicted scanpaths.}
    \label{tab:datasets_related}
\end{table}

\section{Dataset}
\label{sec:dataset}

\paragraph{Prior GUI Datasets}
Existing datasets do not support training models conditioned simultaneously on prompts, wireframes, and visual flows. 
Several datasets have been collected to support GUI-related tasks, as summarized in \autoref{tab:datasets_related}. The AMP dataset~\cite{zhang2021screen} contains 77,000 high-quality screens from 4,068 iOS apps with human annotations, but is not publicly available. On the other hand, the largest publicly available dataset, Rico~\cite{rico}, includes 72,000 app screens from 9,700 Android apps and has been a primary resource for GUI understanding despite its noise issues. To address its limitation, the Clay dataset~\cite{li2022learning} was created by denoising Rico using a pipeline of automated machine learning models and human annotators to provide more accurate element labels. Enrico~\cite{enrico} further cleaned and annotated Rico, but ultimately contains only a small set of high-quality GUIs. MUD~\cite{mud} offers a dataset featuring modern-style Android GUIs. The VINS dataset~\cite{vins} focuses on GUI element detection and was created by manually capturing screenshots from various sources, including both Android and iOS GUIs. Additionally, Webzeitgeist~\cite{kumar2013webzeitgeist} used automated crawling to mine design data from 103,744 webpages, associating web elements with properties such as HTML tags, size, font, and color. Similarly, WebUI~\cite{webui} provides a large-scale collection of website data. However, these datasets neither combine mobile GUIs and webpages nor include visual flow data.
UEyes~\cite{ueyes} is the first mixed GUI-type eye tracking dataset with ground-truth scanpaths, but lacks element labels.

\paragraph{Our Dataset}
To address these limitations, we construct a new large-scale, high-quality dataset combining mobile GUIs and webpages. Our dataset includes approximately 72,500 GUI screenshots, along with their wireframes with labeled GUI elements, descriptions, and scanpaths. This dataset is designed to support the training of generative AI models, filling an important gap in existing public datasets by providing not only GUI images but also detailed descriptions, element labels, and visual flows. 

The dataset includes the following:
\begin{itemize}
\item \textbf{GUI Screenshots}: Our dataset integrates and cleans GUI data from Enrico~\cite{enrico}, VINS~\cite{vins}, and WebUI~\cite{webui}. The Enrico dataset contains 1,460 Android mobile GUIs, while VINS includes 4,543 Android and iOS GUIs. WebUI is a large-scale webpage dataset consisting of approximately 350,000 GUI screenshots with corresponding HTML code. For WebUI, the original dataset includes screenshots for different resolutions, leading to many similar screenshots for each webpage.
We retained only the 1920 x 1080 resolution screenshots to avoid redundant images from different resolutions. We further refined these three datasets by removing abstract, non-graphic wireframe GUIs, duplicates, and GUIs with fewer than three elements.  The final dataset consists of 66,796 webpages and 5,634 mobile GUIs.

\item \textbf{Wireframes with GUI Element Labels}: 
For mobile GUIs, we selected the Enrico and VINS datasets for their well-labeled GUI element bounding boxes. To further refine these annotations, we applied the UIED~\cite{uied} model, which detects and refines GUI element bounding boxes. We manually verified and corrected the results for accuracy. For the WebUI dataset, each element has multiple labels. We filtered the original element labels to keep only the most relevant label for each element. We standardized the labels across mobile and webpage elements, mapping them to nine common types: `Button', `Text', Image', `Icon', `Navigation Bar', `Input Field', `Toggle', `Checkbox', and `Scroll Element'. Using these refined bounding boxes and labels, we generated wireframes with GUI element labels.

\item \textbf{Descriptions}: For each GUI, we applied the LLaVA-Next~\cite{liu2024llavanext} vision-language model to generate both concise and detailed descriptions.

\item \textbf{Scanpaths}: 
To incorporate user attention data, we used EyeFormer~\cite{eyeformer}, the state-of-the-art scanpath prediction model, to predict scanpaths for each GUI. While using real scanpaths recorded by eye trackers would provide more accurate data, this process is highly time-consuming. Alternative proxies, such as webcams or cursor movements, do not capture the same cognitive processes as actual eye movements, making them less suitable for our purposes.
\end{itemize}
\section{Method}
\label{sec:synthesis}


We propose a controllable diffusion-based model \name that enables designers to rapidly explore diverse, realistic GUI designs while flexibly specifying any combination of text prompts, wireframes, and visual flows.
Visual flows are represented as sequences of points, which can be arbitrarily specified. 
In practice, designers typically use them to indicate the desired order in which interface elements should capture user attention. Thus, in this work, we define visual flow as the element order. Our model also supports using a reference GUI, encouraging generated designs to follow the same visual flow (scanpath) as the reference.

Our approach integrates a pretrained diffusion backbone (Stable Diffusion~\cite{rombach2022high}) with two specialized adapters: a \textit{wireframe adapter for local structural control}, aligning generated layouts with wireframes, and a \textit{flow adapter for global perceptual control}, shaping designs to follow desired visual attention patterns. To further enhance realism, we incorporate a \textit{vision–language-guided text correction pipeline} that automatically detects, replaces, and refines illegible or nonsensical placeholder text.

\subsection{Problem Formulation}

We define GUI exploration as a controllable GUI generation task where the model is conditioned on up to three optional modalities: text prompts ($c_\mathrm{t}$), wireframes ($c_\mathrm{w}$), and visual flows ($c_\mathrm{f}$). Text prompts describe style, functionality, or visual attributes in natural language. Wireframes specify spatial layouts and element types. Visual flows define intended user attention sequences, either by providing a reference GUI with the desired flow pattern or by specifying the element order directly. The challenge is to ensure alignment with these conditions while maintaining sufficient diversity for GUI exploration. 

To address this, we employ classifier-free guidance (CFG)~\cite{cfg}, a widely used conditioning technique for diffusion models. CFG works by generating two predictions at each timestep: one with conditioning inputs (e.g., prompt, wireframe, flow) and one without. The final output is an interpolation between them, controlled by a guidance scale parameter. A higher scale enforces stricter adherence to inputs, while a lower scale increases diversity. This balance is essential for design exploration, where both fidelity and variety matter. Without CFG, the model often either ignores inputs or produces repetitive, deterministic outputs.

In addition, we introduce a flow consistency loss $\mathcal{L}_{\mathrm{flow}}$ that encourages alignment between the generated design’s visual flow and the target specification, ensuring user attention is guided as intended without requiring user studies. 
The total objective is:

\begin{equation}
\label{eq:objective_function}
\begin{split}
\mathcal{L}(\hat{z}, c_\mathrm{w}, c_\mathrm{t}, c_\mathrm{f}) = \mathcal{L}_{\mathrm{cfg}}(\hat{z}, c_\mathrm{w}, c_\mathrm{t}, c_\mathrm{f}) +  ~\mathcal{L}_{\mathrm{flow}}(\hat{z}, c_\mathrm{f}),
\end{split}
\end{equation}

where $\hat{z}$ is the generated GUI, $c_\mathrm{w}$, $c_\mathrm{t}$, and $c_\mathrm{f}$ represent the input conditions for the wireframe, text prompt, and visual flow, respectively, and $z_t$ (introduced later) denotes the noisy latent representation at timestep $t$.

\begin{figure}
    \centering
    \includegraphics[width=\linewidth]{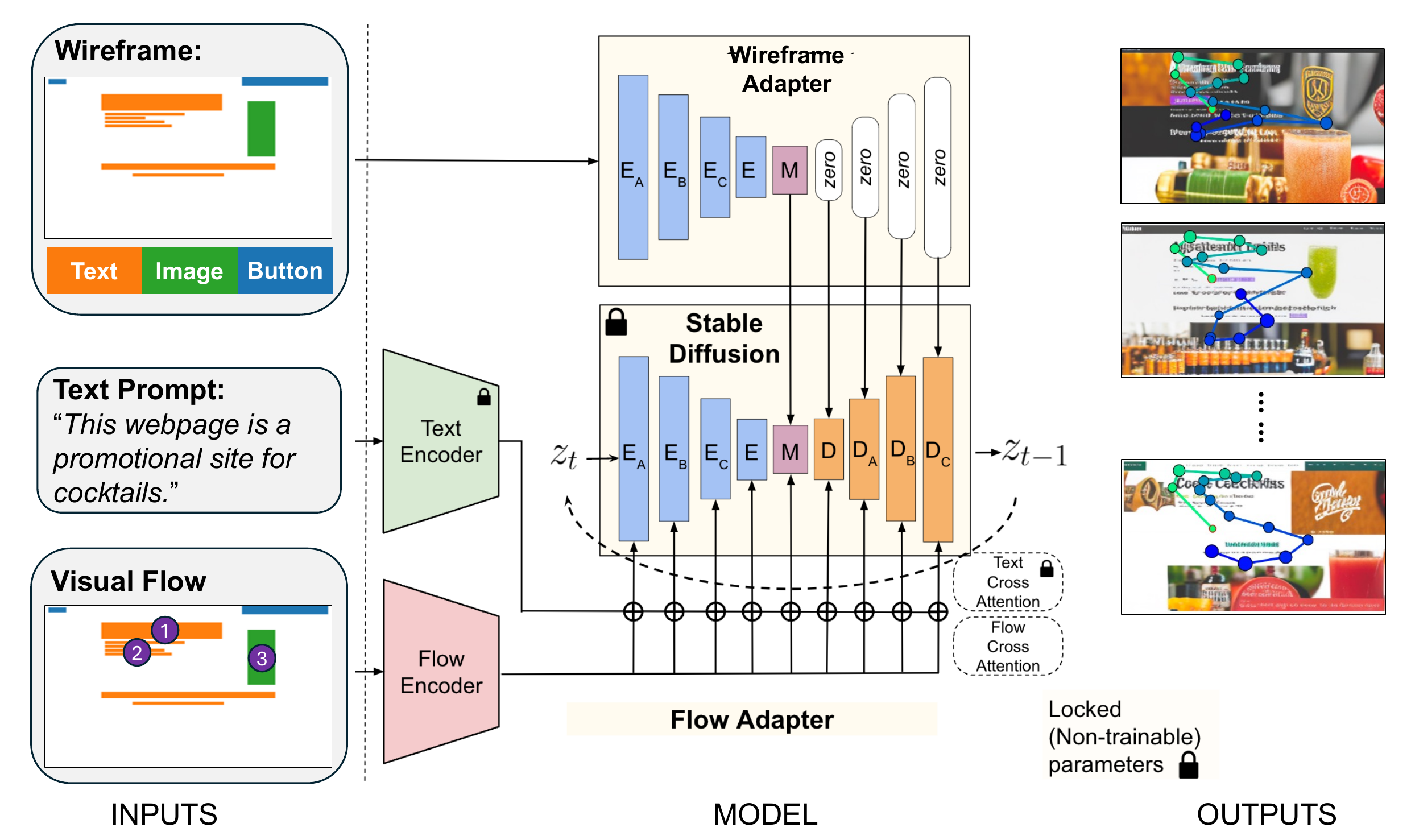}
    \caption{
Our diffusion-based model generates diverse low-fidelity GUIs by conditioning on both local and global properties. The model integrates Stable Diffusion with specialized adapters: the wireframe adapter manages local properties such as element positioning and element types specified on wireframes, while the flow adapter directs the overall visual flow of attention. Given inputs wireframes, prompts, and visual flow patterns, the model effectively produces varied GUI designs.
    }
    \Description{This figure shows that our diffusion-based model generates diverse low-fidelity GUIs by conditioning on both local and global properties. The model integrates Stable Diffusion with specialized adapters: the wireframe adapter manages local properties such as element positioning and element types specified on wireframes, while the flow adapter directs the overall visual flow of attention. Given inputs wireframes, prompts, and visual flow patterns, the model effectively produces varied GUI designs.}
    \label{fig:model}
\end{figure}

\subsection{Controllable GUI Model \name}

We propose a diffusion-based model for GUI generation that conditions on both local properties (e.g., element type and position) and global properties (e.g., visual flow). As illustrated in \autoref{fig:model}, the backbone of our model builds on Stable Diffusion~\cite{rombach2022high}, which employs a U-Net architecture~\cite{ronneberger2015u} consisting of an encoder $E$, a middle block $M$, and a skip-connected decoder $D$. Text prompts are encoded using the CLIP~\cite{clip} text encoder and injected into the diffusion model via cross-attention layers.

\subsubsection{Wireframe Adapter for Local Properties}

Recent advancements in controllable image generation demonstrate that additional networks can be integrated into existing text-to-image diffusion models to improve guidance~\cite{controlnet, controlnet_plus_plus, mou2024t2i}. Inspired by ControlNet~\cite{controlnet}, we construct a trainable copy of Stable Diffusion's encoder and middle block, paired with a decoder augmented with zero-initialized convolution layers. Initializing these convolution weights to zero ensures the adapter only learns deviations from the pretrained backbone. This allows the adapter to efficiently capture local properties, ensuring that the generated GUI aligns with the input wireframe. Wireframe features are concatenated with text embeddings to jointly guide the generation process.

\subsubsection{Flow Adapter for Global Properties}

Cross-attention mechanisms have proven effective in providing global control without explicit spatial guidance~\cite{ye2023ip, zhao2024uni}. We adopt this approach to process visual flow features by adding an additional cross-attention layer in each layer of the diffusion model.
To encode visual flow, we repurpose the decoder of EyeFormer~\cite{eyeformer}, a state-of-the-art scanpath prediction model, to produce embeddings from either a reference GUI or a specified element ordering. These embeddings represent expected user attention trajectories, which then guide the denoising process to respect the intended perceptual flow. This lets designers specify perceptual aspects of the interface without costly user studies.


\subsubsection{Training.}

We train our model using CFG and DDIM (Denoising Diffusion Implicit Models)~\cite{song2020denoising}. DDIM is a deterministic sampler that accelerates denoising while maintaining quality; it produces the same output given the same random seed, ensuring reproducibility for design exploration. Together, CFG and DDIM provide robust control while maintaining efficiency.
Stable Diffusion~\cite{rombach2022high} operates by progressively adding noise to data and learning to reverse this process. 
CFG balances diversity and adherence to input conditions without requiring an additional classifier. Similarly, DDIM deterministically refines noisy latents into clean outputs, producing results that better align with conditioning inputs. More explanation of the sampling process is in the Supplementary Materials.

During training, the Stable Diffusion backbone remains frozen, while our adapters are optimized. For a given timestep $t$ and input conditions $C = {c_\mathrm{w}, c_\mathrm{t}, c_\mathrm{f}}$, the model learns to predict the added noise $\epsilon_{\theta}$:

\begin{equation}
    \mathcal{L}_{\textrm{cfg}}(\hat{z}, C) = \mathbb{E}_{\hat{z}, t, C, \epsilon \sim \mathcal{N}(0,1)} [\| \epsilon - \epsilon_{\theta}(z_t, t, \mathcal{C}) \|^{2}_2]
\end{equation}

where $z_t$ is the noisy latent at timestep $t$, and $\hat{z} = z_0$ is the final denoised GUI.

We train each adapter independently. Specifically, the wireframe adapter is trained using both text prompt and wireframe inputs, while the flow adapter is trained using text prompts and visual flow inputs. During training, we apply a 50‡

\paragraph{Loss Function for Wireframe Adapter} 

The wireframe adapter is trained solely with the CFG loss:
\begin{equation}
  \mathcal{L_\textrm{WireframeAdapter}} =  \mathcal{L}_{\textrm{cfg}}(\hat{z}, c_w, c_t) = \mathbb{E}_{z_t, t, c_w, c_t \epsilon \sim \mathcal{N}(0,1)} [\| \epsilon - \epsilon_{\theta}(z_t, t, c_w, c_t) \|^{2}_2]
\end{equation}

\paragraph{Loss Function for Flow Adapter} 
The flow adapter is trained using both CFG loss and a flow consistency loss. The latter ensures that the cross-attention layers guide generation toward a latent subspace aligned with the desired visual flow.

To encourage the consistency between the visual flow of the generated GUIs and the input visual flow specification, we apply Dynamic Time Warping (DTW)~\cite{dtw}, a standard metric used to measure the similarity between two temporal sequences, even when they differ in length. DTW identifies the optimal match between the sequences and computes the distance between them. However, since the original DTW is not differentiable, it cannot be directly used to optimize deep learning models.
To address this, we employ softDTW~\cite{soft-dtw}, a differentiable version of DTW suitable for deep learning.

Since the only output of the network is a synthesized GUI ($\hat{z}$) with no ground-truth scanpath, we use EyeFormer~\cite{eyeformer} to compute a representative ground truth ($\textrm{EyeFormer}(\hat{z}) \sim \hat{c}_v$). The loss is defined as:
\begin{equation}
    \mathcal{L}_{\textrm{flow}}(\hat{z}, c_f) = \textrm{softDTW}(\textrm{EyeFormer}(\hat{z}, c_f))
\end{equation}

Thus, the total loss for the flow adapter is:
\begin{equation}
\begin{split}
  \mathcal{L_\textrm{FlowAdapter}} &=  \mathcal{L}_{\textrm{cfg}}(\hat{z}, c_t, c_f) + \mathcal{L}_{\textrm{flow}}(\hat{z}, c_f) \\
  &= \mathbb{E}_{z_t, t, c_t, c_f \epsilon \sim \mathcal{N}(0,1)} [\| \epsilon - \epsilon_{\theta}(z_t, t,c_t, c_f) \|^{2}_2] + \textrm{softDTW}(\textrm{EyeFormer}(\hat{z}, c_f))).
  \end{split}
\end{equation}

\subsection{Implementation Details.}
We train at a resolution of 256x256 on our combined dataset of 72,500+ real GUIs ranging from web pages to mobile apps (\autoref{tab:datasets_related}). Text prompts are automatically generated using Llava~\cite{liu2023llava}, wireframes are obtained from labeled datasets, and visual flow conditions are generated on the fly using EyeFormer~\cite{eyeformer}.
We train our framework in 3 independent stages: First, we train the wireframe adapter using prompts and wireframes. Then we train the flow adapter using prompts and visual flows. Finally, we jointly finetune with all three inputs (wireframe, text, visual flow).
We train stage 1 for 10 epochs and stage 2 for 350k iterations. All experiments use full precision training (to avoid instability in adapters), AdamW~\cite{adamW} with learning rate $1e^{-4}$, and weight decay 0.01.
Experiments run on 2 NVIDIA GeForce RTX 4090 GPUs (24 GB VRAM each). Our implementation uses the HuggingFace Diffusers library~\cite{diffusers_lib} and PyTorch~\cite{pytorch}. 


\section{EVALUATION METHOD}
\label{sec:experiments}

We evaluate \name through qualitative and quantitative evaluations.
Our experiments address two main research questions:
\begin{enumerate}
    \item Can \name generate diverse GUIs more aligned with text, wireframe, and visual flow inputs compared to other baseline models?
    \item How does multimodal conditioning affect the generated GUIs' perceptual quality, prompt alignment, and flow alignment?
\end{enumerate}

\subsection{Dataset}
We evaluate on our new multimodal GUI dataset introduced in Section~\ref{sec:dataset}, which contains 72,500 GUIs spanning both mobile screens and webpages. Each GUI includes screenshots, wireframes with labeled elements, textual descriptions, and predicted scanpaths. For evaluation, we split the dataset into 80\% training and 20\% testing, ensuring no app or website overlaps between splits.

\subsection{Baseline Models}

No prior model can generate GUIs that explicitly incorporate visual flow specifications.
Thus, we compare \name against four representative baseline models that can generate GUIs based on text prompts and wireframes.
\begin{itemize}
 \item \textbf{ControlNet}~\cite{controlnet}: A diffusion-based controllable generation model that integrates structural conditions such as edges and poses. We adapt it to wireframe-based GUI generation.
\item \textbf{IP-Adapter}~\cite{ye2023ip}: A cross-attention–based adapter for Stable Diffusion that incorporates additional input modalities, enabling image- and layout-conditioned generation.
\item \textbf{Stitch}~\cite{stitch}: A GUI-specific generative model that synthesizes screen layouts through hierarchical recombination of interface components.
\item \textbf{GPT-5}: A Large Language Model (LLM) with multimodal capabilities that can take text prompts and wireframes as input to generate GUIs.
\end{itemize}

\subsection{Metrics}
\label{sec:metrics}

We evaluate generated GUIs using complementary metrics that capture perceptual quality, text prompt alignment, and visual flow alignment:

\begin{itemize}
\item \textbf{ManIQA}\cite{yang2022maniqa}. A non-reference perceptual quality metric designed to detect complex and non-traditional distortions, penalizing unrealistic generations.
\item \textbf{MUSIQ}\cite{musiq}. A transformer-based multi-resolution metric that assigns scores based on perceptual quality across image scales.
\item \textbf{CLIP-IQA}\cite{clip}. A zero-shot metric that uses CLIP embeddings to measure semantic alignment between a generated GUI and its textual description.
\item \textbf{softDTW}. A differentiable dynamic time warping metric that measures similarity between predicted and target scanpaths. We generate pseudo ground-truth scanpaths from real GUIs using EyeFormer~\cite{eyeformer} and compare them against scanpaths extracted from generated GUIs.
\end{itemize}

ManIQA and MUSIQ together assess relative perceptual quality. CLIP-IQA complements them by evaluating semantic alignment with prompts, while softDTW measures how well generated GUIs guide user attention in the intended order.

\section{Results}

\begin{figure*}[!]
 \def\w{\linewidth}
 \centering
  \includegraphics[width=0.8\w]{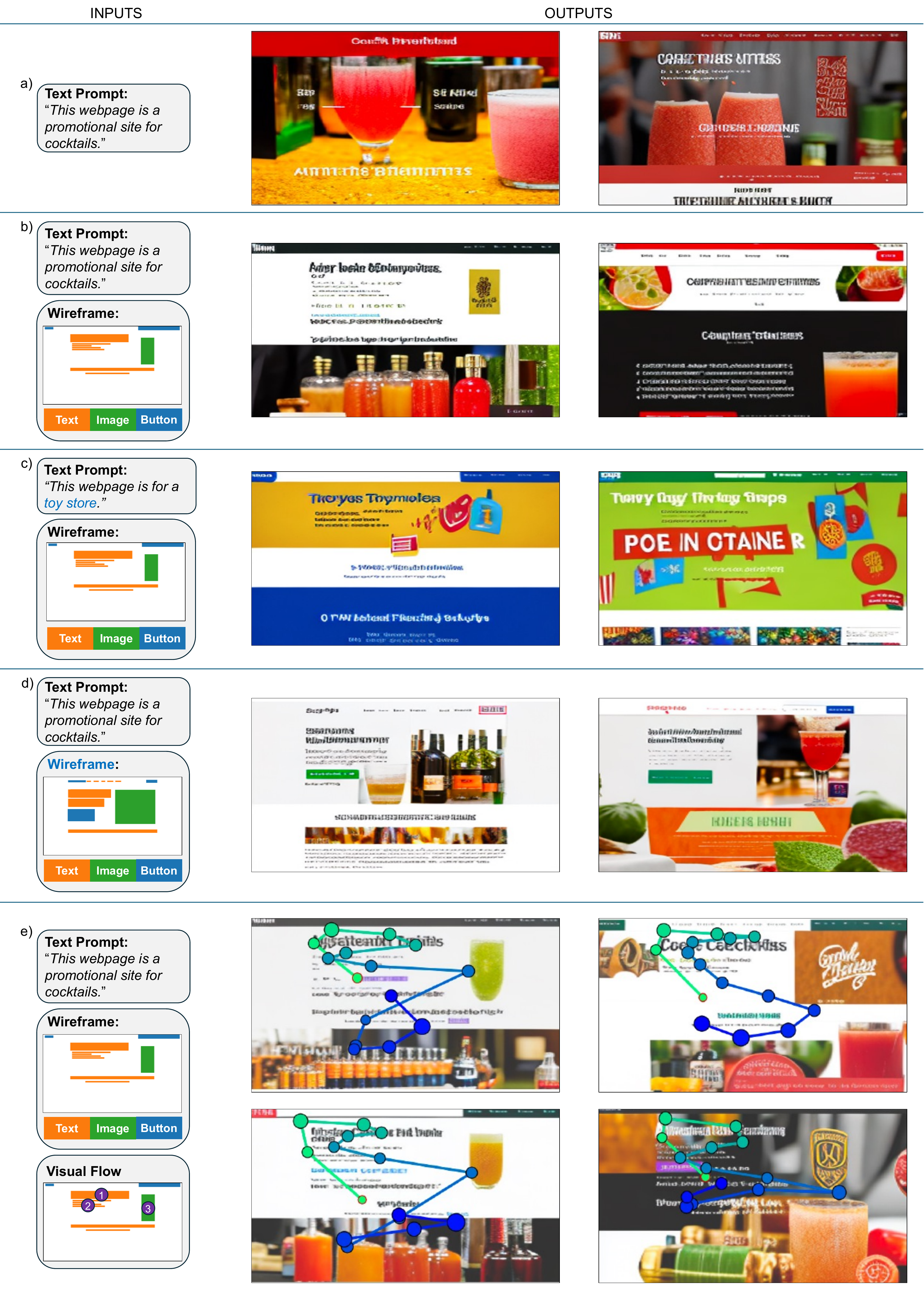}
\caption{
\name has the capability of generating diverse GUIs under different input conditions. a) With only a text prompt, the model generates diverse GUIs consistent with the given description. b,c,d) Adding a wireframe enforces structural fidelity, ensuring elements such as text, images, and buttons follow the specified layout. Results show that varying the wireframe or text prompt leads to distinct GUI topic and structural variations. e) Incorporating visual flow further constrains generation by guiding user attention across elements in the intended order. Scanpaths are visualized using a color gradient (green → blue) indicating temporal progression, with circles marking fixation points, closely matching the specified sequence.
}
\Description{This figure shows the qualitative results under different input conditions. a) With only a text prompt, the model generates diverse GUIs consistent with the given description. b,c,d) Adding a wireframe enforces structural fidelity, ensuring elements such as text, images, and buttons follow the specified layout. Results show that varying the wireframe or text prompt leads to distinct GUI topic and structural variations. e) Incorporating visual flow further constrains generation by guiding user attention across elements in the intended order.}
\label{fig:result1}
\end{figure*}

\begin{figure*}[!]
 \def\w{\linewidth}
 \centering
  \includegraphics[width=0.83\w]{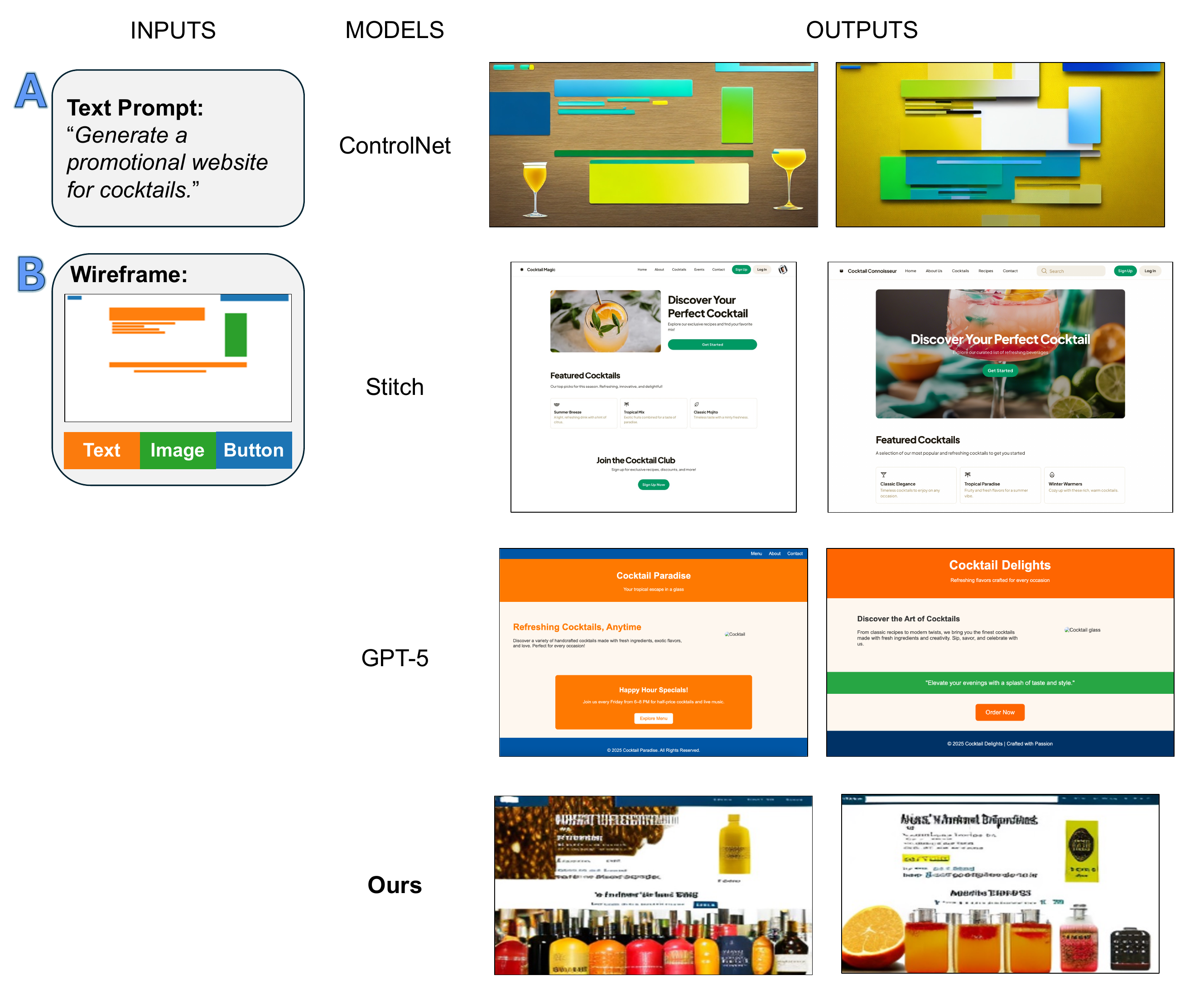}
\caption{
Comparison of GUI generation results across baseline models and \name. Given the same text prompt and wireframe, ControlNet produces abstract layouts without meaningful semantic alignment. Stitch enforces structural coherence but results in overly simplistic and template-like designs. GPT-5 generates textually coherent websites, but layouts are rigid and lack stylistic diversity. In contrast, our model \name produces realistic, visually rich GUIs that simultaneously adhere to the wireframe structure, reflect the semantics of the prompt, and provide stylistic diversity suitable for early design exploration.
}
\Description{This figure shows the comparison of GUI generation results across baseline models and ControlGUI.}
\label{fig:Sept2025_comparison}
\end{figure*}

We evaluate \name through qualitative and quantitative studies, spanning both mobile GUIs and webpages, and testing across all input modality combinations.

We first present qualitative results to demonstrate controllability and comparisons against representative baselines.
We then report quantitative results across perceptual, semantic, and flow metrics, followed by an analysis of computational efficiency.

\subsection{Qualitative Evaluation}

\autoref{fig:Sept2025_comparison} illustrates how \name produces diverse and controllable GUI designs across different input modalities. With only text prompts, the model captures high-level semantics (e.g., ``a promotional site for cocktails'') while producing stylistically varied layouts. Incorporating wireframes enforces structural consistency, aligning generated GUIs with element placement and type. We observe that even small changes in wireframes lead to substantial variations in layout while maintaining semantic coherence with the prompt. Finally, when visual flows are provided, generated GUIs not only respect textual and structural constraints but also exhibit attention trajectories that follow the desired sequence. This demonstrates the model’s ability to jointly align with semantic, structural, and perceptual specifications.
More results are shown in the Supplementary Materials.

Beyond demonstrating controllability with different input modalities, we also qualitatively compare \name against representative baselines: ControlNet, Stitch, and GPT-5 (\autoref{fig:Sept2025_comparison}). ControlNet excels at preserving geometric layouts but fails to capture semantic fidelity, often producing abstract colored blocks that resemble placeholders rather than usable GUIs. Stitch generates structurally coherent outputs with recognizable UI components, but its designs are largely template-driven and lack stylistic variation, which limits their value for open-ended ideation. GPT-5 produces websites that are textually coherent and consistent with the given prompt, yet its layouts tend to be rigid and oversimplified, offering little flexibility in supporting wireframe-driven design. In contrast, \name respects structural constraints from wireframes, aligns with semantic details from text prompts, and produces results with varied aesthetics. These comparisons underscore the key advantage of \name: its ability to unify semantic, structural, and perceptual conditions in a single generative process, providing designers with realistic yet flexible GUI concepts that better support rapid ideation and broad exploration of alternatives.

\begin{table*}[t]
  \caption{Quantitative comparison between our model and baselines. \name outperforms ControlNet and IP-Adapter in perceptual quality (ManIQA, MUSIQ) and prompt alignment (CLIP-IQA), while uniquely supporting flow alignment (Soft-DTW), which baselines cannot model.}
  \Description{ This table shows the quantitative comparison between our model and baselines.}
  \label{tab:quantitative_comparison}
  \begin{tabular}{lcccc}
    \toprule
    \multirow{2}{*}{\textbf{Method}} & 
    \multicolumn{2}{c}{\textbf {Perceptual Quality}} & \multicolumn{1}{c}{\textbf {Prompt Alignment}}  & \multicolumn{1}{c}{\textbf {Flow Alignment}} \\
    \cmidrule(lr){2-3} 
    \cmidrule(lr){4-4} 
    \cmidrule(lr){5-5} 
     & ManIQA~\cite{yang2022maniqa} $\uparrow$ & MUSIQ~\cite{musiq} $\uparrow$ &  ClipIQA~\cite{clipIQA} $\uparrow$  & Soft-DTW~\cite{soft-dtw} $\downarrow$\\      
    \midrule

    ControlNet~\cite{controlnet} & 0.39 & 41.22 & 0.43 & NA \\  
    IP-Adapter~\cite{ye2023ip}  & 0.52 & 60.34 & 0.51 & NA \\
    \midrule
    \textbf{Ours}  & \bf 0.61 &  \bf60.44 &  \bf0.53 &  \bf0.55 \\
    \bottomrule
  \end{tabular}
\end{table*}

\begin{table*}[t]
  \caption{Different input modality combinations. Each variant (V1–V7) corresponds to a subset of modalities: wireframe, flow, and/or prompts. Results reveal trade-offs: using fewer modalities often yields slightly higher perceptual quality, while incorporating flow improves alignment of attention sequences (lower Soft-DTW).}
  \Description{This table shows different input modality combinations. Each variant (V1–V7) corresponds to a subset of modalities: wireframe, flow, and/or prompts. Results reveal trade-offs: using fewer modalities often yields slightly higher perceptual quality, while incorporating flow improves alignment of attention sequences (lower Soft-DTW).}
  \label{tab:ablation_inputs}
  \resizebox{\linewidth}{!}{
  \begin{tabular}{lccccccc}
  \toprule
  \multirow{2}{*}{\textbf{Methods}} & \multicolumn{3}{c}{\textbf {Inputs}} & \multicolumn{2}{c}{\textbf {Perceptual Quality}} & \multicolumn{1}{c}{\textbf {Prompt Alignment}}  & \multicolumn{1}{c}{\textbf {Flow Alignment}} \\      
  \cmidrule(lr){2-4} 
  \cmidrule(lr){5-7}
  \cmidrule(lr){8-8}
  & Wireframe & Flow & Prompts & ManIQA~\cite{yang2022maniqa} $\uparrow$  &  MUSIQ~\cite{musiq} $\uparrow$ & ClipIQA~\cite{clipIQA} $\uparrow$  & Soft-DTW~\cite{soft-dtw} $\downarrow$  \\      
  \midrule
    V1 & \cmark & \xmark  & \xmark  & 0.66 &  64.73 & 0.48 & NA  \\  
    V2  & \xmark & \cmark & \xmark & 0.62 & 62.03  & 0.53 & 0.36 \\   
    V3 & \xmark & \xmark & \cmark & 0.66 & 67.05  & 0.53 & NA \\   
    V4 & \cmark & \cmark & \xmark & 0.60 & 61.63 & 0.50 & 0.32 \\   
    V5 & \cmark & \xmark & \cmark & 0.63 & 61.73 & 0.57 & NA \\   
    V6 & \xmark & \cmark & \cmark & 0.60 & 59.11 & 0.56 & 0.34 \\   
    V7 & \cmark & \cmark & \cmark & 0.59 & 59.63 & 0.54 & 0.31 \\
    \bottomrule
  \end{tabular}
  }
\end{table*}

\subsection{Quantitative Evaluation}

We quantitatively evaluate \name through two sets of experiments: (1) comparison against strong baselines, and (2) ablation studies over different input modality combinations.

\paragraph{Baseline Comparison.}
\autoref{tab:quantitative_comparison} reports results against ControlNet and IP-Adapter on 1,000 randomly sampled test cases. We measure perceptual realism using ManIQA and MUSIQ, prompt alignment using CLIP-IQA, and flow alignment using Soft-DTW. \name achieves the strongest perceptual quality (0.61 ManIQA, 60.44 MUSIQ), surpassing both baselines, and also shows superior prompt alignment (0.53 CLIP-IQA). More importantly, \name is the only method that supports visual flow inputs, yielding a Soft-DTW score of 0.55. Neither ControlNet nor IP-Adapter can explicitly model user attention sequences. This comparison highlights the unique advantage of \name’s dual-adapter design, which enables controllable synthesis across multiple modalities.

\paragraph{Different input modality combinations.}

Our model can input any combination of the three input types. \autoref{tab:ablation_inputs}  explores how different combinations of text prompts, wireframes, and flows affect performance. Because our model supports multimodal conditioning, we evaluate the same \name model under all seven possible combinations of inputs.

The results reveal consistent trade-offs across modalities. Variants without flow inputs (V1, V3, V5) achieve slightly higher perceptual quality, since the model faces fewer constraints and can optimize for visual smoothness. However, these settings cannot guide user attention, which limits their utility for designers concerned with perceptual experience. Conversely, adding flow inputs (V2, V4, V6, V7) introduces mild drops in perceptual scores but enables explicit control over attention sequences. 

Wireframes also have a distinct effect. Variants that exclude wireframes (V2, V6) show reasonable prompt alignment but reduced structural fidelity, producing layouts that look visually appealing but may place elements inconsistently. By contrast, variants including wireframes (V3, V4, V5, V7) preserve element positions and types more faithfully, albeit with a small trade-off in perceptual quality due to stricter constraints.

\paragraph{Efficiency}

In terms of runtime, \name generates a batch of 16 GUI candidates in approximately 19 seconds on one RTX 4090 GPU. GUIs within a single batch often share stylistic similarity, while those across different batches display greater diversity. This efficiency makes the model practical for iterative design exploration, where designers benefit from quickly exploring multiple candidate GUIs.
\section{User Study}

To evaluate how our controllable GUI exploration model supports early-stage GUI ideation, we conducted a within-subjects user study. The study investigates whether our model enables designers to generate diverse interface sketches with minimal effort, and how varying levels of input control (prompt, wireframe, visual flow) affect perceived control, intent alignment, and satisfaction. 
The aim was to assess both the impact on design efficiency and the subjective experience. 

The study was designed with four primary objectives: (1) determine whether \name helps users generate more diverse interface ideas with less effort;
(2) assess how users perceive control, satisfaction, and intent alignment across input modalities, compared to a prompt-only baseline;
(3) investigate whether multimodal inputs enhance creativity beyond baseline results; and
(4) identify limitations of \name and potential directions for improvement.

\subsection{Study Design}

\subsubsection{Participants}

We recruited 18 participants (8 males, 10 females, ages 23–30, SD=2.48) through email and social media. All participation was anonymized. All the participants have design experience or have used AI models to generate GUI design. Each session lasted approximately 60 minutes, and participants received 16€ in compensation. The study was approved through local ethics procedures.

\subsubsection{Materials}
The study was conducted through a Gradio-based web interface where participants could provide text prompts, wireframes, and element orderings as input to our model. GUIs were generated and displayed as a gallery for comparison. Questionnaires were collected through Google Forms on participants’ personal devices.

\subsubsection{Experiment Design}
A within-subjects design was employed, where each participant experienced both generating a GUI with a prompt only and with additional wireframes and/or the order of elements.

\subsubsection{Procedure}

After providing consent and demographic information, participants completed a short tutorial introducing the system. They were then assigned three design tasks in counterbalanced order, including: (1) designing a homepage for a personal finance dashboard (with the wireframe provided), and (2) designing a cocktail promotion webpage (with the provided wireframe and flow order for the first three fixations). Finally, participants designed one GUI of their choice.
After completing the tasks, participants filled out a short questionnaire comparing the results generated from prompt-only inputs (baseline) with those produced using multimodal inputs. At the end of the study, participants also completed a brief open-ended interview to provide overall impressions, comment on the generated results, and offer suggestions for improvement.

\subsection{Perceived Diversity of the Designs}

\begin{figure*}[!]
 \def\w{\linewidth}
 \centering
  \includegraphics[width=0.83\w]{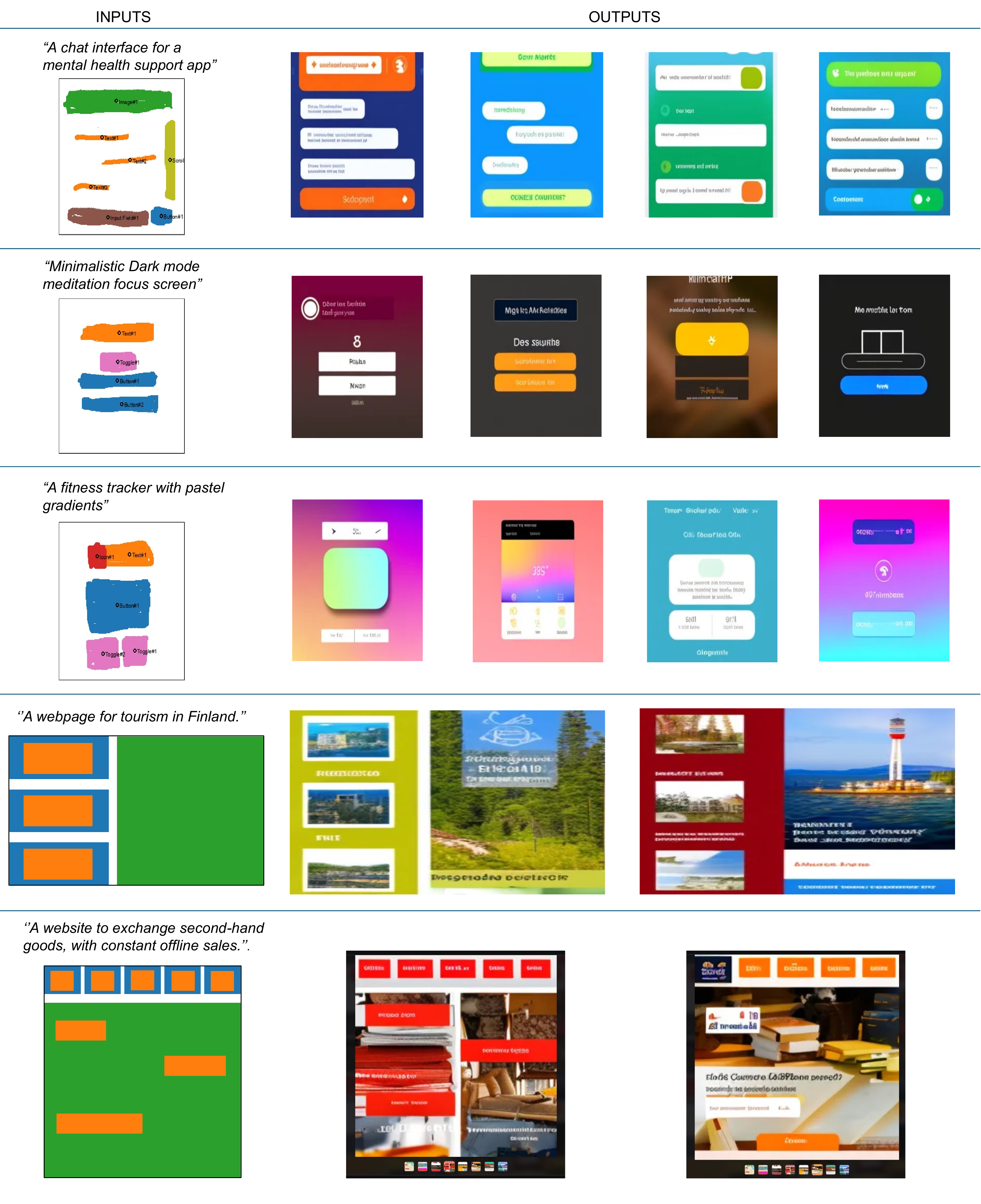}
\caption{Examples of GUIs generated by participants during the user study. Each row illustrates a different design brief (left: text prompt + wireframe input, right: generated outputs). The examples span a range of application scenarios. Despite being trained on wireframes with rigid bounding boxes, the model is also able to handle hand-drawn wireframes, producing diverse and design variants aligned with user intent.
}
\Description{This figure shows examples of GUIs generated by participants during the user study. Each row illustrates a different design brief (left: text prompt + wireframe input, right: generated outputs). The examples span a range of application scenarios. Despite being trained on wireframes with rigid bounding boxes, the model is also able to handle hand-drawn wireframes, producing diverse and design variants aligned with user intent.}
\label{fig:Sept2025_user}
\end{figure*}

\begin{figure*}[!]
 \def\w{\linewidth}
 \centering
  \includegraphics[width=\w]{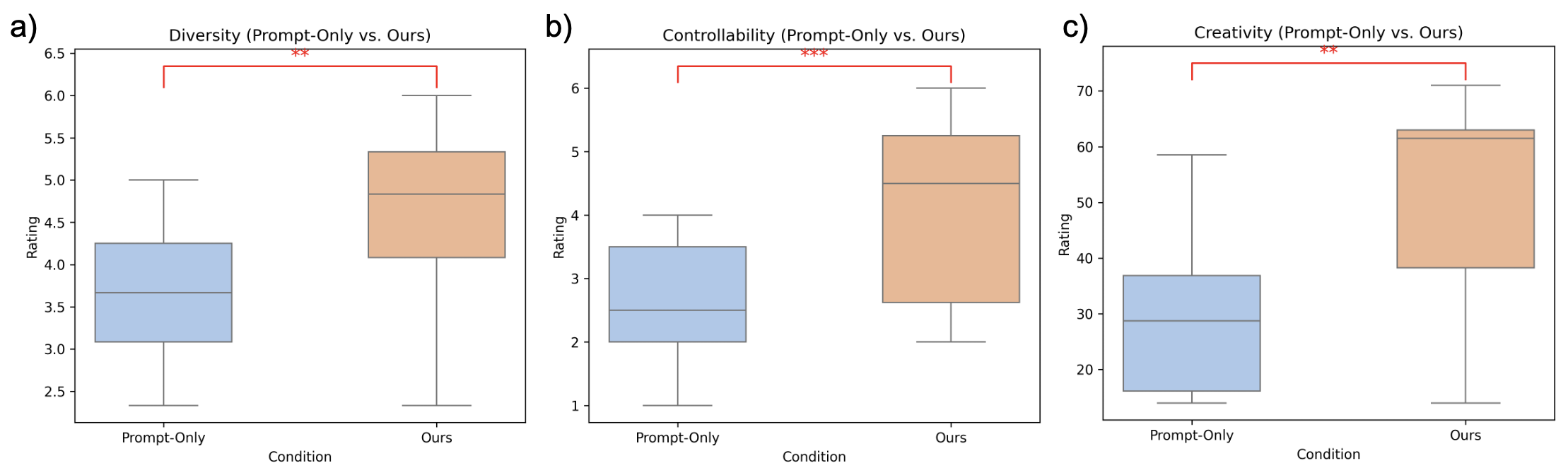}
\caption{Participants’ subjective ratings of diversity, controllability, and creativity when comparing prompt-only input (baseline) against our multimodal system. Across both dimensions, participants rated our system significantly higher, showing the benefits of combining prompts with wireframes and visual flows to guide GUI generation.
}
\Description{This figure shows participants’ subjective ratings of diversity, controllability, and creativity when comparing prompt-only input (baseline) against our multimodal system. Across both dimensions, participants rated our system significantly higher, showing the benefits of combining prompts with wireframes and visual flows to guide GUI generation.}
\label{fig:Sept2025_plot}
\end{figure*}

\autoref{fig:Sept2025_user} shows some samples of the GUIs participants generated for different tasks. More results are shown in the Supplementary Materials. Although the model was trained on wireframes with rigid bounding boxes, our model can also generate diverse GUI results given hand-drawn wireframes.
The results in \autoref{fig:user_ratings}a show that participants rated our system significantly higher in terms of diversity than the prompt-only baseline.
On a 7-point Likert scale, the average diversity rating for \name was 4.59, compared to 3.56 for the prompt-only condition. This difference was statistically significant (Wilcoxon test: $W = 1.5$, $p = 0.0021 < 0.01$). 

Qualitative feedback further highlights the impact of diversity. Participants emphasized how \name expanded their design space and revealed unexpected ideas. For example, some participants mentioned that ``I think the diversity of the results is good, in terms of colors, layouts, and images'' (P3) and ``Your system provides several fancy examples that I could refer to'' (P14).
For many participants, this breadth was central to ideation. For example, P8 stated ``It inspires me to continue iterating on my thoughts'', while P1 explained, ``It extends my imagination about image choices and color matching''. 

Participants also valued being able to compare multiple alternatives at once. As P3 remarked, ``I could compare diverse generated examples, maybe I can combine several parts I like together'' and P14 also mentioned ``I really like the system because I can try out my ideas and explore as many trials as I want''. 

Even when outputs were imperfect, participants still appreciated the diversity they provided. As P15 observed, ``although the generated photos are not always normal, they offer me a way to think about new ideas.'' 
Taken together, both ratings and participant reflections demonstrate that \name effectively broadens the design space, supports divergent thinking, and inspires creative exploration.

\subsection{Perceived Control, Satisfaction, Alignment with Intent}

As shown in \autoref{fig:Sept2025_plot}b, participants rated our system significantly higher in terms of controllability compared to the prompt-only baseline. On a 7-point Likert scale, the average rating for \name was 4.00, versus 2.58 for prompt-only input. It shows a statistically significant improvement (Wilcoxon test: $W = 0.0$, $p = 0.0004 < 0.001$).

Qualitative feedback reinforces this result. Participants consistently emphasized that wireframes and visual flows gave them stronger control over the generation process and helped align outputs with their design intent. P16 explained ``Your system offers me the control over the design with the wireframe input. It is a good improvement based on the baseline''.
Others highlighted the practicality of wireframes for specifying layout, e.g., ``It is more efficient to talk to the machine about what kind of component I mean to use in a low-fi wireframe'' (P8). and ``Your system is more efficient for those who have basic design experience but are not professionals, because they have thought about the general structure of the website, but not the contents inside'' (P2).

Participants also noted that controllability translated into stronger alignment between outputs and their intent. For example, P6 commented ``the generated results are more consistent with the inputs''. 

In sum, the findings show that multimodal inputs not only improved participants’ sense of control but also enhanced satisfaction by producing designs that better align with their desired outcomes. Compared to prompt-only generation, enabling more modalities provided a more intuitive and structured way to meet design intent.

\subsection{Creativity}

We measured perceived creativity using the Creative Support Index (CSI). On average, participants rated \name substantially higher (51.31) compared to the prompt-only baseline (31.11), a statistically significant improvement (Wilcoxon test: $W = 5.0$, $p = 0.0011$). While the absolute CSI scores were moderate likely due to the low-fidelity of current GUI generations, the relative gain demonstrates that multimodal control still enhanced participants’ creative experience.

In the user study, Participants frequently described the system as a catalyst for creativity, particularly in the early stages of ideation. The diversity of outputs encouraged them to think beyond their initial ideas, e.g., ``It inspires me to continue iterating my thoughts'' (P8), ``It extends my imagination about the photo choice direction and color matching'' (P1), and ``I could also compare diverse generated examples, combining several parts I like together. There are also creative designs I had not come up with'' (P3).

Beyond inspiration, participants described the system as useful for validating and refining ideas. P11 commented ``The system will help the designers to validate the design ideas'', and P14 stressed ``I really like the system because I could try my ideas and explore as many trials as I could''.

In summary, the findings show that multimodal control not only enhanced creativity scores quantitatively but also expanded participants’ creative space qualitatively. By surfacing unexpected and diverse examples, the system supported divergent thinking and validation during early-stage ideation.

\subsection{Limitations of \name}

While participants appreciated the controllability and diversity offered by \name, they also highlighted several limitations that affect its practicality for design work.

A recurring concern was output quality. Many noted that the generated GUIs were often low-resolution, with unreadable or unstable text. P4 mentioned ``low resolution in generated images and text, unstable generations according to prompt''. Others agreed that ``the quality of the generated texts and images is not plausible'' (P2) and ``the running speed is a little low'' (P3). 

Finally, participants reflected on the professional applicability of the \name, e.g., ``Considering the quality of the UI output, honestly speaking, I work much faster and more efficient without the tool. The current output level is some process an average UX/UI designer can have inside their brain already'' (P8) and ``I am working for complex B2B enterprise software solutions … the tool cannot understand professional use case and scenario very well'' (P8).






\section{Discussion} 

\begin{figure*}
    \centering
    \includegraphics[width=0.7\linewidth]{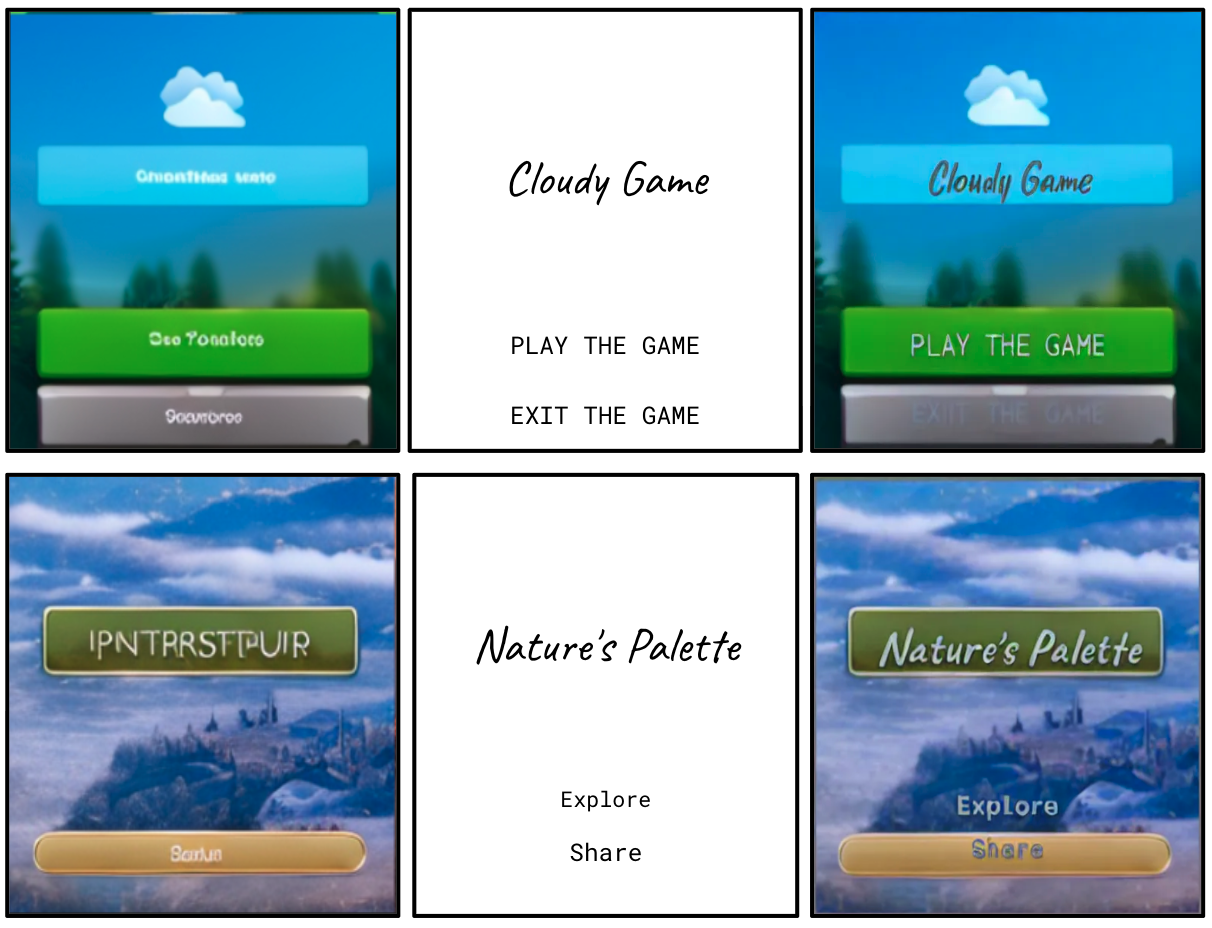}
    \caption{Example results for text enhancement. The left column shows raw GUI generations with illegible or unrealistic text. The middle column shows the corrected text content proposed by our VLM-guided text correction pipeline. The right column presents the final enhanced GUIs with legible and context-appropriate text integrated into the original design.}
    \Description{This figure shows example results for text enhancement. The left column shows raw GUI generations with illegible or unrealistic text. The middle column shows the corrected text content proposed by our VLM-guided text correction pipeline. The right column presents the final enhanced GUIs with legible and context-appropriate text integrated into the original design.}
    \label{fig:text_correction}
\end{figure*}

Our results suggest that diffusion-based models are suitable for the rapid and flexible exploration of low-fidelity GUI generation.
Currently, no available diffusion model is capable of producing GUIs with high-quality text and graphics, but this limitation does not impede their use for the exploration of early-stage design ideas.
There the focus is more on generating a broad range of 'good enough' possibilities.
Specifically, we demonstrated that the \emph{probabilistic} nature of diffusion models allows for generating a multitude of diverse GUI ideas efficiently. 

Before this paper, it was an open question of how to design generative models such that designers do not need to over-specify the input.
By extending conditional diffusion to handle both wireframes and visual flow, we offer a low-effort approach to designers that simplifies interaction with diffusion models. 
Moreover, the inclusion of specialized adapters for considering local (e.g., the position and type of GUI elements) and global (e.g., visual flow) design properties enhances control over GUI characteristics, allowing designers to focus on broad exploration with minimal effort. 
We view this model as a step forward in creating user-adapted GUI designs with generated models, as it integrates GUI properties with user-centered interactions, such as eye movement, throughout the GUI design exploration process.

\subsection{Text Enhancement}

A recurring limitation identified in our user study is the ``text problem'' in diffusion-based GUI generation. Participants frequently complained that generated GUIs contained unreadable text, gibberish placeholders, or inconsistent typography, which reduced realism and made the outputs harder to use for actual ideation. 

To address this challenge, we further propose a vision–language–guided text correction pipeline that automatically detects illegible or placeholder text and replaces it with context-appropriate alternatives. This process requires no additional effort from the designer, enabling more realistic and semantically coherent prototypes while maintaining rapid exploration.

Technically, we build on TextDiffuser~\cite{textdiffuser}, an inpainting-based diffusion model for text rendering.
To provide contextually appropriate replacements, we employ LLaVA-Next~\cite{liu2024llavanext}, a high-resolution vision–language model trained for fine-grained multimodal reasoning. 
We use the model in single-shot mode, combining the user’s original generation prompt with a structured system prompt:
\noindent\texttt{
<Prepended User-Prompt>. This is shown in the image. Give a list of answers in the format: []\textbackslash n Text at $(x_1, y_1)$: <text-at-$(x_1,y_1)$>\textbackslash n … }
This yields automatically localized and meaningful text suggestions for each region. 
 Our approach avoids interrupting the rapid prototyping cycle and preserves the efficiency of GUI generation.
The initial results are shown in~\autoref{fig:text_correction}.

\subsection{Limitations \& Future Work} 

We acknowledge the following limitations. 
First, we face challenges in generating realistic human faces, with some outputs appearing distorted. 
Increasing the dataset of human faces could improve the model's training. 
Second, the generated resolution is low since we trained on 256x256-resolution images. 
Future work could benefit from better computing power to train on higher-resolution images. 
Third, our model currently does not support highly specific requirements, such as generating a Margarita cocktail on a website. 
Additionally, our visual flow results lack finer control. 
We used the model EyeFormer~\cite{eyeformer} for scanpath prediction, however, it often biases scanpaths, starting from the center and moving toward the top left and right, limiting control over specific element emphasis. Although these patterns reflect natural eye movements, GUI designers may require more flexibility to direct user attention. 
Lastly, our current method does not support iterative design, which is important for early-stage prototyping, as each generation results in entirely new GUIs. 
Future work could improve the diffusion model by incorporating conditioning on the outputs from previous iterations. 
This could be done by using the current design as input to generate subsequent iterations. 
By conditioning on previous results, the model could produce variations that build upon existing designs, rather than starting anew each time.

\section{Conclusion}

We presented \name, a controllable diffusion model that introduces multimodal inputs into early-stage GUI exploration. Our results demonstrate that moving beyond prompt-only generation substantially improves designers’ ability to explore the design space. Compared to text-only systems, \name enables the rapid synthesis of diverse alternatives that better align with both structural and perceptual intent. This breadth not only increases diversity but also acts as a catalyst for creativity, helping designers extend their imagination, explore directions they might not otherwise consider, and iterate more effectively on early concepts.

A central reason for these gains lies in the dual-adapter diffusion architecture, which simultaneously models local structure through wireframe constraints and global perceptual goals through visual flow. This design allows \name to generate outputs that are not only realistic in layout but also aligned with intended attention sequences, which is something prior systems could not achieve. By allowing designers to specify any combination of text, wireframe, and flow inputs, \name provides flexible multimodal control that adapts to natural design practices rather than forcing rigid input formats.

Together, these capabilities make \name well-suited for supporting early-stage ideation, where the goal is to explore many low-fidelity yet meaningful possibilities before committing to detailed implementation. While current diffusion models still face challenges in generating high-quality text and polished visuals, our study shows that even imperfect outputs can meaningfully support divergent thinking, idea validation, and rapid exploration.

\bibliographystyle{ACM-Reference-Format}
\bibliography{Reference}

@inproceedings{herring2009getting,
  title={Getting inspired! Understanding how and why examples are used in creative design practice},
  author={Herring, Scarlett R and Chang, Chia-Chen and Krantzler, Jesse and Bailey, Brian P},
  booktitle={Proceedings of the SIGCHI conference on human factors in computing systems},
  pages={87--96},
  year={2009}
}

@misc{stitch,
  title = {Stitch},
  howpublished = {\url{https://stitch.withgoogle.com/}},
  note = {Accessed: 2015-09-01}
}

@misc{liu2024llavanext,
    title={LLaVA-NeXT: Improved reasoning, OCR, and world knowledge},
    url={https://llava-vl.github.io/blog/2024-01-30-llava-next/},
    author={Liu, Haotian and Li, Chunyuan and Li, Yuheng and Li, Bo and Zhang, Yuanhan and Shen, Sheng and Lee, Yong Jae},
    month={January},
    year={2024}
}

@misc{liu2023llava,
      title={Visual Instruction Tuning}, 
      author={Liu, Haotian and Li, Chunyuan and Wu, Qingyang and Lee, Yong Jae},
      publisher={NeurIPS},
      year={2023},
}

@inproceedings{li2022learning,
  title={Learning to denoise raw mobile UI layouts for improving datasets at scale},
  author={Li, Gang and Baechler, Gilles and Tragut, Manuel and Li, Yang},
  booktitle={Proceedings of the 2022 CHI Conference on Human Factors in Computing Systems},
  pages={1--13},
  year={2022}
}

@inproceedings{swearngin2020scout,
  title={Scout: Rapid exploration of interface layout alternatives through high-level design constraints},
  author={Swearngin, Amanda and Wang, Chenglong and Oleson, Alannah and Fogarty, James and Ko, Amy J},
  booktitle={Proceedings of the 2020 CHI conference on human factors in computing systems},
  pages={1--13},
  year={2020}
}

@article{wu2024uicoder,
  title={UICoder: Finetuning Large Language Models to Generate User Interface Code through Automated Feedback},
  author={Wu, Jason and Schoop, Eldon and Leung, Alan and Barik, Titus and Bigham, Jeffrey P and Nichols, Jeffrey},
  journal={arXiv preprint arXiv:2406.07739},
  year={2024}
}

@inproceedings{li2021screen2vec,
  title={Screen2vec: Semantic embedding of gui screens and gui components},
  author={Li, Toby Jia-Jun and Popowski, Lindsay and Mitchell, Tom and Myers, Brad A},
  booktitle={Proceedings of the 2021 CHI Conference on Human Factors in Computing Systems},
  pages={1--15},
  year={2021}
}

@inproceedings{kumar2013webzeitgeist,
  title={Webzeitgeist: design mining the web},
  author={Kumar, Ranjitha and Satyanarayan, Arvind and Torres, Cesar and Lim, Maxine and Ahmad, Salman and Klemmer, Scott R and Talton, Jerry O},
  booktitle={Proceedings of the SIGCHI Conference on Human Factors in Computing Systems},
  pages={3083--3092},
  year={2013}
}

@article{jansson1991design,
  title={Design fixation},
  author={Jansson, David G and Smith, Steven M},
  journal={Design studies},
  volume={12},
  number={1},
  pages={3--11},
  year={1991},
  publisher={Elsevier}
}

@inproceedings{tohidi2006getting,
  title={Getting the right design and the design right},
  author={Tohidi, Maryam and Buxton, William and Baecker, Ronald and Sellen, Abigail},
  booktitle={Proceedings of the SIGCHI conference on Human Factors in computing systems},
  pages={1243--1252},
  year={2006}
}

@inproceedings{dow2011prototyping,
  title={Prototyping dynamics: sharing multiple designs improves exploration, group rapport, and results},
  author={Dow, Steven and Fortuna, Julie and Schwartz, Dan and Altringer, Beth and Schwartz, Daniel and Klemmer, Scott},
  booktitle={Proceedings of the SIGCHI conference on human factors in computing systems},
  pages={2807--2816},
  year={2011}
}

@inproceedings{landay1996silk,
  title={SILK: sketching interfaces like krazy},
  author={Landay, James A},
  booktitle={Conference companion on Human factors in computing systems},
  pages={398--399},
  year={1996}
}

@article{boyarski1994computers,
  title={Computers and communication design: Exploring the rhetoric of HCI},
  author={Boyarski, Daniel and Buchanan, Richard},
  journal={Interactions},
  volume={1},
  number={2},
  pages={25--35},
  year={1994},
  publisher={ACM New York, NY, USA}
}

@inproceedings{ueyes,
	author = {Jiang, Yue and Leiva, Luis A. and Rezazadegan Tavakoli, Hamed and R. B. Houssel, Paul and Kylm\"{a}l\"{a}, Julia and Oulasvirta, Antti},
	title = {UEyes: Understanding Visual Saliency across User Interface Types},
	year = {2023},
	isbn = {9781450394215},
	publisher = {Association for Computing Machinery},
	address = {New York, NY, USA},
	url = {https://doi.org/10.1145/3544548.3581096},
	doi = {10.1145/3544548.3581096},
	booktitle = {Proceedings of the 2023 CHI Conference on Human Factors in Computing Systems},
	articleno = {285},
	numpages = {21},
	location = {Hamburg, Germany},
	series = {CHI '23}
}

@article{cheng2023play,
  title={Play: Parametrically conditioned layout generation using latent diffusion},
  author={Cheng, Chin-Yi and Huang, Forrest and Li, Gang and Li, Yang},
  journal={arXiv preprint arXiv:2301.11529},
  year={2023}
}

@misc{clip,
      title={Learning Transferable Visual Models From Natural Language Supervision}, 
      author={Alec Radford and Jong Wook Kim and Chris Hallacy and Aditya Ramesh and Gabriel Goh and Sandhini Agarwal and Girish Sastry and Amanda Askell and Pamela Mishkin and Jack Clark and Gretchen Krueger and Ilya Sutskever},
      year={2021},
      eprint={2103.00020},
      archivePrefix={arXiv},
      primaryClass={cs.CV},
      url={https://arxiv.org/abs/2103.00020}, 
}

@article{cheng2024colay,
  title={CoLay: Controllable Layout Generation through Multi-conditional Latent Diffusion},
  author={Cheng, Chin-Yi and Gao, Ruiqi and Huang, Forrest and Li, Yang},
  journal={arXiv preprint arXiv:2405.13045},
  year={2024}
}

@article{Rosenholtz11,
 author = {Rosenholtz, Ruth and Dorai, Amal and Freeman, Rosalind},
 title = {Do Predictions of Visual Perception Aid Design?},
 journal = {ACM Trans. Appl. Percept.},
 volume = {8},
 number = {2},
 year = {2011},
 _pages = {12:1--12:20},
}

@inproceedings{Still10,
  author = {Jeremiah D. Still and Christopher M. Masciocchi },
  title = {A Saliency Model Predicts Fixations in Web Interfaces},
  booktitle = {Proc. MDDAUI Workshop},
  year = {2010},
  pages = {},
}

@misc{side_tune,
      title={Side-Tuning: A Baseline for Network Adaptation via Additive Side Networks}, 
      author={Jeffrey O Zhang and Alexander Sax and Amir Zamir and Leonidas Guibas and Jitendra Malik},
      year={2020},
      eprint={1912.13503},
      archivePrefix={arXiv},
      primaryClass={cs.LG},
      url={https://arxiv.org/abs/1912.13503}, 
}

@InProceedings{adapters_in_nlp,
  title = 	 {Parameter-Efficient Transfer Learning for {NLP}},
  author =       {Houlsby, Neil and Giurgiu, Andrei and Jastrzebski, Stanislaw and Morrone, Bruna and De Laroussilhe, Quentin and Gesmundo, Andrea and Attariyan, Mona and Gelly, Sylvain},
  booktitle = 	 {Proceedings of the 36th International Conference on Machine Learning},
  pages = 	 {2790--2799},
  year = 	 {2019},
  editor = 	 {Chaudhuri, Kamalika and Salakhutdinov, Ruslan},
  volume = 	 {97},
  series = 	 {Proceedings of Machine Learning Research},
  month = 	 {09--15 Jun},
  publisher =    {PMLR},
  url = 	 {https://proceedings.mlr.press/v97/houlsby19a.html}
}

@misc{eyeformer,
    title={EyeFormer: Predicting Personalized Scanpaths with Transformer-Guided Reinforcement Learning},
    author={Jiang, Yue and Guo, Zixin and Tavakoli, Hamed Rezazadegan and Leiva, Luis A and Oulasvirta, Antti},
    publisher={arXiv},
    eprint={2404.10163},
    url={https://arxiv.org/abs/2404.10163},
    year={2024},
}

@article{rico_semantic,
  author    = {Srinivas Sunkara and
               Maria Wang and
               Lijuan Liu and
               Gilles Baechler and
               Yu-Chung Hsiao and
               Jindong Chen and
               Abhanshu Sharma and
               James Stout},
  title     = {Towards Better Semantic Understanding of Mobile Interfaces},
  journal   = {CoRR},
  volume    = {abs/2210.02663},
  year      = {2022},
  url       = {https://arxiv.org/abs/2210.02663},
  eprinttype = {arXiv},
  eprint    = {2210.02663},
}

@inproceedings{mud,
author = {Feng, Sidong and Ma, Suyu and Wang, Han and Kong, David and Chen, Chunyang},
title = {MUD: Towards a Large-Scale and Noise-Filtered UI Dataset for Modern Style UI Modeling},
year = {2024},
isbn = {9798400703300},
publisher = {Association for Computing Machinery},
address = {New York, NY, USA},
url = {https://doi.org/10.1145/3613904.3642350},
doi = {10.1145/3613904.3642350},
booktitle = {Proceedings of the CHI Conference on Human Factors in Computing Systems},
articleno = {7},
numpages = {14},
location = {Honolulu, HI, USA},
series = {CHI '24}
}

@inproceedings{rico,
author = {Deka, Biplab and Huang, Zifeng and Franzen, Chad and Hibschman, Joshua and Afergan, Daniel and Li, Yang and Nichols, Jeffrey and Kumar, Ranjitha},
title = {Rico: A Mobile App Dataset for Building Data-Driven Design Applications},
year = {2017},
isbn = {9781450349819},
publisher = {Association for Computing Machinery},
address = {New York, NY, USA},
url = {https://doi.org/10.1145/3126594.3126651},
doi = {10.1145/3126594.3126651},
booktitle = {Proceedings of the 30th Annual ACM Symposium on User Interface Software and Technology},
pages = {845–854},
numpages = {10},
location = {Qu\'{e}bec City, QC, Canada},
series = {UIST '17}
}

@inproceedings{jiang2019orclayout,
author = {Jiang, Yue and Du, Ruofei and Lutteroth, Christof and Stuerzlinger, Wolfgang},
title = {ORC Layout: Adaptive GUI Layout with OR-Constraints},
year = {2019},
isbn = {9781450359702},
publisher = {Association for Computing Machinery},
address = {New York, NY, USA},
url = {https://doi.org/10.1145/3290605.3300643},
doi = {10.1145/3290605.3300643},
booktitle = {Proceedings of the 2019 CHI Conference on Human Factors in Computing Systems},
articleno = {Paper 413},
numpages = {12},
location = {Glasgow, Scotland Uk},
series = {CHI ’19}
}

@inproceedings{jiang2020orcsolver,
author = {Jiang, Yue and Stuerzlinger, Wolfgang and Zwicker, Matthias and Lutteroth, Christof},
title = {ORCSolver: An Efficient Solver for Adaptive GUI Layout with OR-Constraints},
year = {2020},
isbn = {9781450367080},
publisher = {Association for Computing Machinery},
address = {New York, NY, USA},
url = {https://doi.org/10.1145/3313831.3376610},
doi = {10.1145/3313831.3376610},
booktitle = {Proceedings of the 2020 CHI Conference on Human Factors in Computing Systems},
pages = {1–14},
numpages = {14},
location = {Honolulu, HI, USA},
series = {CHI ’20}
}

@inproceedings{jiang2020reverseorc,
author = {Jiang, Yue and Stuerzlinger, Wolfgang and Lutteroth, Christof},
title = {ReverseORC: Reverse Engineering of Resizable User Interface Layouts with OR-Constraints},
year = {2021},
isbn = {9781450380966},
publisher = {Association for Computing Machinery},
address = {New York, NY, USA},
url = {https://doi.org/10.1145/3411764.3445043},
doi = {10.1145/3411764.3445043},
booktitle = {Proceedings of the 2021 CHI Conference on Human Factors in Computing Systems},
articleno = {316},
numpages = {18},
location = {Yokohama, Japan},
series = {CHI '21}
}

@inproceedings{jiang2022computational,
author = {Jiang, Yue and Lu, Yuwen and Nichols, Jeffrey and Stuerzlinger, Wolfgang and Yu, Chun and Lutteroth, Christof and Li, Yang and Kumar, Ranjitha and Li, Toby Jia-Jun},
title = {Computational Approaches for Understanding, Generating, and Adapting User Interfaces},
year = {2022},
isbn = {9781450391566},
publisher = {Association for Computing Machinery},
address = {New York, NY, USA},
url = {https://doi.org/10.1145/3491101.3504030},
doi = {10.1145/3491101.3504030},
booktitle = {Extended Abstracts of the 2022 CHI Conference on Human Factors in Computing Systems},
articleno = {74},
numpages = {6},
location = {New Orleans, LA, USA},
series = {CHI EA '22}
}

@inproceedings{jiang2023future,
author = {Jiang, Yue and Lu, Yuwen and Lutteroth, Christof and Li, Toby Jia-Jun and Nichols, Jeffrey and Stuerzlinger, Wolfgang},
title = {The Future of Computational Approaches for Understanding and Adapting User Interfaces},
year = {2023},
isbn = {9781450394222},
publisher = {Association for Computing Machinery},
address = {New York, NY, USA},
url = {https://doi.org/10.1145/3544549.3573805},
doi = {10.1145/3544549.3573805},
booktitle = {Extended Abstracts of the 2023 CHI Conference on Human Factors in Computing Systems},
articleno = {367},
numpages = {5},
location = {Hamburg, Germany},
series = {CHI EA '23}
}

@inproceedings{jiang2024computational,
  title={Computational Methodologies for Understanding, Automating, and Evaluating User Interfaces},
  author={Jiang, Yue and Lu, Yuwen and Kliman-Silver, Clara and Lutteroth, Christof and Li, Toby Jia-Jun and Nichols, Jeffrey and Stuerzlinger, Wolfgang},
  booktitle={Extended Abstracts of the 2024 CHI Conference on Human Factors in Computing Systems},
  year={2024},
series = {CHI EA '24}
}

@inproceedings{zhang2021screen,
author = {Zhang, Xiaoyi and de Greef, Lilian and Swearngin, Amanda and White, Samuel and Murray, Kyle and Yu, Lisa and Shan, Qi and Nichols, Jeffrey and Wu, Jason and Fleizach, Chris and Everitt, Aaron and Bigham, Jeffrey P},
title = {Screen Recognition: Creating Accessibility Metadata for Mobile Applications from Pixels},
year = {2021},
isbn = {9781450380966},
publisher = {Association for Computing Machinery},
address = {New York, NY, USA},
url = {https://doi.org/10.1145/3411764.3445186},
doi = {10.1145/3411764.3445186},
booktitle = {Proceedings of the 2021 CHI Conference on Human Factors in Computing Systems},
articleno = {275},
numpages = {15},
location = {Yokohama, Japan},
series = {CHI '21}
}

@inproceedings{jiang2024flexdoc,
  title={FlexDoc: Flexible Document Adaptation through Optimizing both Content and Layout},
  author={Jiang, Yue and Lutteroth, Christof and Jain, Rajiv and Tensmeyer, Christopher and Manjunatha, Varun and Stuerzlinger, Wolfgang and Morariu, Vlad I},
  booktitle={2024 IEEE Symposium on Visual Languages and Human-Centric Computing (VL/HCC)},
  pages={217--222},
  year={2024},
  organization={IEEE}
}

@article{emami2024impact,
  title={Impact of Design Decisions in Scanpath Modeling},
  author={Emami, Parvin and Jiang, Yue and Guo, Zixin and Leiva, Luis A},
  journal={Proceedings of the ACM on Human-Computer Interaction},
  volume={8},
  number={ETRA},
  pages={1--16},
  year={2024},
  publisher={ACM New York, NY, USA}
}

@article{wang2024visrecall++,
  title={VisRecall++: Analysing and Predicting Visualisation Recallability from Gaze Behaviour},
  author={Wang, Yao and Jiang, Yue and Hu, Zhiming and Ruhdorfer, Constantin and B{\^a}ce, Mihai and Bulling, Andreas},
  journal={Proceedings of the ACM on Human-Computer Interaction},
  volume={8},
  number={ETRA},
  pages={1--18},
  year={2024},
  publisher={ACM New York, NY, USA}
}

@article{jiang2024ueyes,
  title={UEyes: An Eye-Tracking Dataset across User Interface Types},
  author={Jiang, Yue and Leiva, Luis A and Houssel, Paul RB and Tavakoli, Hamed R and Kylm{\"a}l{\"a}, Julia and Oulasvirta, Antti},
  journal={arXiv preprint arXiv:2402.05202},
  year={2024}
}

@inproceedings{jiang2024computational2,
  title={Computational Representations for Graphical User Interfaces},
  author={Jiang, Yue},
  booktitle={Extended Abstracts of the 2024 CHI Conference on Human Factors in Computing Systems},
  year={2024},
series = {CHI EA '24}
}

@inproceedings{vins,
author = {Bunian, Sara and Li, Kai and Jemmali, Chaima and Harteveld, Casper and Fu, Yun and Seif El-Nasr, Magy Seif},
title = {VINS: Visual Search for Mobile User Interface Design},
year = {2021},
isbn = {9781450380966},
publisher = {Association for Computing Machinery},
address = {New York, NY, USA},
url = {https://doi.org/10.1145/3411764.3445762},
doi = {10.1145/3411764.3445762},
booktitle = {Proceedings of the 2021 CHI Conference on Human Factors in Computing Systems},
articleno = {423},
numpages = {14},
location = {Yokohama, Japan},
series = {CHI '21}
}

@inproceedings{
adamW,
title={Decoupled Weight Decay Regularization},
author={Ilya Loshchilov and Frank Hutter},
booktitle={International Conference on Learning Representations},
year={2019},
url={https://openreview.net/forum?id=Bkg6RiCqY7},
}

@article{textdiffuser,
        title={TextDiffuser: Diffusion Models as Text Painters},
        author={Chen, Jingye and Huang, Yupan and Lv, Tengchao and Cui, Lei and Chen, Qifeng and Wei, Furu},
        journal={arXiv preprint arXiv:2305.10855},
        year={2023}
      }

@article{ye2023ip,
  title={Ip-adapter: Text compatible image prompt adapter for text-to-image diffusion models},
  author={Ye, Hu and Zhang, Jun and Liu, Sibo and Han, Xiao and Yang, Wei},
  journal={arXiv preprint arXiv:2308.06721},
  year={2023}
}

@incollection{pytorch,
title = {PyTorch: An Imperative Style, High-Performance Deep Learning Library},
author = {Paszke, Adam and Gross, Sam and Massa, Francisco and Lerer, Adam and Bradbury, James and Chanan, Gregory and Killeen, Trevor and Lin, Zeming and Gimelshein, Natalia and Antiga, Luca and Desmaison, Alban and Kopf, Andreas and Yang, Edward and DeVito, Zachary and Raison, Martin and Tejani, Alykhan and Chilamkurthy, Sasank and Steiner, Benoit and Fang, Lu and Bai, Junjie and Chintala, Soumith},
booktitle = {Advances in Neural Information Processing Systems 32},
pages = {8024--8035},
year = {2019},
publisher = {Curran Associates, Inc.},
url = {http://papers.neurips.cc/paper/9015-pytorch-an-imperative-style-high-performance-deep-learning-library.pdf}
}

@article{song2020denoising,
  title={Denoising diffusion implicit models},
  author={Song, Jiaming and Meng, Chenlin and Ermon, Stefano},
  journal={arXiv preprint arXiv:2010.02502},
  year={2020}
}

@software{diffusers_lib,
author = {von Platen, Patrick and Patil, Suraj and Lozhkov, Anton and Cuenca, Pedro and Lambert, Nathan and Rasul, Kashif and Davaadorj, Mishig and Nair, Dhruv and Paul, Sayak and Liu, Steven and Berman, William and Xu, Yiyi and Wolf, Thomas},
license = {Apache-2.0},
title = {{Diffusers: State-of-the-art diffusion models}},
url = {https://github.com/huggingface/diffusers},
version = {0.12.1}
}

@article{zhao2024uni,
  title={Uni-controlnet: All-in-one control to text-to-image diffusion models},
  author={Zhao, Shihao and Chen, Dongdong and Chen, Yen-Chun and Bao, Jianmin and Hao, Shaozhe and Yuan, Lu and Wong, Kwan-Yee K},
  journal={Advances in Neural Information Processing Systems},
  volume={36},
  year={2024}
}

@inproceedings{ronneberger2015u,
  title={U-net: Convolutional networks for biomedical image segmentation},
  author={Ronneberger, Olaf and Fischer, Philipp and Brox, Thomas},
  booktitle={Medical image computing and computer-assisted intervention--MICCAI 2015: 18th international conference, Munich, Germany, October 5-9, 2015, proceedings, part III 18},
  pages={234--241},
  year={2015},
  organization={Springer}
}

@inproceedings{rombach2022high,
  title={High-resolution image synthesis with latent diffusion models},
  author={Rombach, Robin and Blattmann, Andreas and Lorenz, Dominik and Esser, Patrick and Ommer, Bj{\"o}rn},
  booktitle={Proceedings of the IEEE/CVF conference on computer vision and pattern recognition},
  pages={10684--10695},
  year={2022}
}

@inproceedings{mou2024t2i,
  title={T2i-adapter: Learning adapters to dig out more controllable ability for text-to-image diffusion models},
  author={Mou, Chong and Wang, Xintao and Xie, Liangbin and Wu, Yanze and Zhang, Jian and Qi, Zhongang and Shan, Ying},
  booktitle={Proceedings of the AAAI Conference on Artificial Intelligence},
  volume={38},
  number={5},
  pages={4296--4304},
  year={2024}
}

@article{saharia2022photorealistic,
  title={Photorealistic text-to-image diffusion models with deep language understanding},
  author={Saharia, Chitwan and Chan, William and Saxena, Saurabh and Li, Lala and Whang, Jay and Denton, Emily L and Ghasemipour, Kamyar and Gontijo Lopes, Raphael and Karagol Ayan, Burcu and Salimans, Tim and others},
  journal={Advances in neural information processing systems},
  volume={35},
  pages={36479--36494},
  year={2022}
}

@article{ramesh2022hierarchical,
  title={Hierarchical text-conditional image generation with clip latents},
  author={Ramesh, Aditya and Dhariwal, Prafulla and Nichol, Alex and Chu, Casey and Chen, Mark},
  journal={arXiv preprint arXiv:2204.06125},
  volume={1},
  number={2},
  pages={3},
  year={2022}
}

@article{nichol2021glide,
  title={Glide: Towards photorealistic image generation and editing with text-guided diffusion models},
  author={Nichol, Alex and Dhariwal, Prafulla and Ramesh, Aditya and Shyam, Pranav and Mishkin, Pamela and McGrew, Bob and Sutskever, Ilya and Chen, Mark},
  journal={arXiv preprint arXiv:2112.10741},
  year={2021}
}

@article{ho2022classifier,
  title={Classifier-free diffusion guidance},
  author={Ho, Jonathan and Salimans, Tim},
  journal={arXiv preprint arXiv:2207.12598},
  year={2022}
}

@article{dhariwal2021diffusion,
  title={Diffusion models beat gans on image synthesis},
  author={Dhariwal, Prafulla and Nichol, Alexander},
  journal={Advances in neural information processing systems},
  volume={34},
  pages={8780--8794},
  year={2021}
}

@inproceedings{uied,
author = {Xie, Mulong and Feng, Sidong and Xing, Zhenchang and Chen, Jieshan and Chen, Chunyang},
title = {UIED: a hybrid tool for GUI element detection},
year = {2020},
isbn = {9781450370431},
publisher = {Association for Computing Machinery},
address = {New York, NY, USA},
url = {https://doi.org/10.1145/3368089.3417940},
doi = {10.1145/3368089.3417940},
booktitle = {Proceedings of the 28th ACM Joint Meeting on European Software Engineering Conference and Symposium on the Foundations of Software Engineering},
pages = {1655–1659},
numpages = {5},
location = {Virtual Event, USA},
series = {ESEC/FSE 2020}
}

@InProceedings{enrico,
  author    = {Luis A. Leiva and Asutosh Hota and Antti Oulasvirta},
  title     = {Enrico: A High-quality Dataset for Topic Modeling of Mobile {UI} Designs},
  booktitle = {Proc. MobileHCI Adjunct},
  year      = {2020},
}

@misc{cfg,
      title={Classifier-Free Diffusion Guidance}, 
      author={Jonathan Ho and Tim Salimans},
      year={2022},
      eprint={2207.12598},
      archivePrefix={arXiv},
      primaryClass={cs.LG},
      url={https://arxiv.org/abs/2207.12598}, 
}

@misc{controlnet,
  title={Adding Conditional Control to Text-to-Image Diffusion Models}, 
  author={Lvmin Zhang and Anyi Rao and Maneesh Agrawala},
  booktitle={IEEE International Conference on Computer Vision (ICCV)},
  year={2023}
}

@article{webui, 
    title={WebUI: A Dataset for Enhancing Visual UI Understanding with Web Semantics}, 
    author={Jason Wu and Siyan Wang and Siman Shen and Yi-Hao Peng and Jeffrey Nichols and Jeffrey Bigham}, 
    journal={ACM Conference on Human Factors in Computing Systems (CHI)}, 
    year={2023}
}

@inproceedings{controlnet_plus_plus,
  title={ControlNet++: Improving Conditional Controls with Efficient Consistency Feedback: Project Page: liming-ai. github. io/ControlNet\_Plus\_Plus},
  author={Li, Ming and Yang, Taojiannan and Kuang, Huafeng and Wu, Jie and Wang, Zhaoning and Xiao, Xuefeng and Chen, Chen},
  booktitle={European Conference on Computer Vision},
  pages={129--147},
  year={2024},
  organization={Springer}
}

@ARTICLE{dtw,
  author={Sakoe, H. and Chiba, S.},
  journal={IEEE Transactions on Acoustics, Speech, and Signal Processing}, 
  title={Dynamic programming algorithm optimization for spoken word recognition}, 
  year={1978},
  volume={26},
  number={1},
  pages={43-49},
  doi={10.1109/TASSP.1978.1163055}}

@inproceedings{soft-dtw,
author = {Cuturi, Marco and Blondel, Mathieu},
title = {Soft-DTW: a differentiable loss function for time-series},
year = {2017},
publisher = {JMLR.org},
booktitle = {Proceedings of the 34th International Conference on Machine Learning - Volume 70},
pages = {894–903},
numpages = {10},
location = {Sydney, NSW, Australia},
series = {ICML'17}
}

@inproceedings{yang2022maniqa,
  title={MANIQA: Multi-dimension Attention Network for No-Reference Image Quality Assessment},
  author={Yang, Sidi and Wu, Tianhe and Shi, Shuwei and Lao, Shanshan and Gong, Yuan and Cao, Mingdeng and Wang, Jiahao and Yang, Yujiu},
  booktitle={Proceedings of the IEEE/CVF Conference on Computer Vision and Pattern Recognition},
  pages={1191--1200},
  year={2022}
}

@inproceedings{clipIQA,
    author = {Wang, Jianyi and Chan, Kelvin CK and Loy, Chen Change},
    title = {Exploring CLIP for Assessing the Look and Feel of Images},
    booktitle = {AAAI},
    year = {2023}
}

@InProceedings{musiq,
    author    = {Ke, Junjie and Wang, Qifei and Wang, Yilin and Milanfar, Peyman and Yang, Feng},
    title     = {MUSIQ: Multi-Scale Image Quality Transformer},
    booktitle = {Proceedings of the IEEE/CVF International Conference on Computer Vision (ICCV)},
    month     = {October},
    year      = {2021},
    pages     = {5148-5157}
}

\end{document}